\documentclass[prd,twocolumn,showpacs,preprintnumbers,nofootinbib]{revtex4}
\usepackage[dvips]{graphicx}
\usepackage{enumerate}
\usepackage{amsmath,amssymb}
\usepackage{mathrsfs}
\usepackage[dvips]{graphicx}
\usepackage{bm}

\begin{document}
\preprint{CHIBA-EP-202-v3, 2013}

\title{
Stability of chromomagnetic condensation 
and 
mass generation 
for confinement in SU(2) Yang-Mills theory
}

\author{Kei-Ichi Kondo$^{1}$}
\email{kondok@faculty.chiba-u.jp}

\affiliation{$^1$Department of Physics,  
Graduate School of Science, 
Chiba University, Chiba 263-8522, Japan
}
\begin{abstract}

We show that the Nielsen-Olesen instability of the Savvidy vacuum with a homogeneous chromomagnetic condensation  disappears in the framework of the functional renormalization group. 
This result follows from our observations: 
(i) the vanishing imaginary part of the  effective average action is realized for arbitrary infrared cutoff as a novel fixed point solution of the  flow equation for the complex-valued effective average action 
and  
(ii)  an approximate analytical solution for the  effective average action is obtained without the pure imaginary part for large infrared cutoff.  
This result suggests that there exists   a physical mechanism for maintaining the stability or staying on the fixed point even for sufficiently small infrared cutoff. 
We argue that dynamical gluon mass  generation (related to two-gluon bound states identified  with glueballs) occurs due to the Becchi-Rouet-Stora-Tyutin-invariant vacuum condensate of mass dimension two without causing    instability.

\end{abstract}


\maketitle
%%%%%%%%%%%%%%%%%%%%%%%%%%%%%%%%%%%%%%%%%%%%%%%%%

\section{Introduction}

%%%%%%%%%%%%%%%%%%%%%%%%%%%%%%%%%%%%%%%%%%%%%%%%%

The \textbf{dual superconductor picture} for the Yang-Mills theory  \cite{YM54} vacuum is an attractive hypothesis for explaining quark confinement. 
It has been intensively investigated  as a promising mechanism for quark confinement  up to today since the early proposal in the 1970s by Nambu, Mandelstam, 't Hooft and Polyakov 
\cite{dualsuper}. 
In an ordinary (type II) superconductor, electric charges condense into \textbf{Cooper pairs}. As a result magnetic flux is squeezed into tubes. 
In the dual superconductor picture of the Yang-Mills theory vacuum, \textbf{chromomagnetic monopoles} are to be condensed into dual Cooper pairs and the \textbf{chromoelectric flux} connecting  color charges  is to be squeezed into tubes forming the \textbf{hadron string}. 
 Then the nonzero string tension plays the role of the constant of proportionality in the linear potential realizing quark confinement. 
The key ingredients of this picture are the existence of \textbf{chromomagnetic monopole condensation} and the \textbf{dual Meissner effect}.

There are two methods available to define the  chromomagnetic monopole  in the Yang-Mills theory:
\begin{enumerate}
\item
  \textbf{Abelian projection}  (by ' t Hooft \cite{tHooft81}):

partial gauge fixing of the gauge group $G$ to the maximal torus subgroup: $G \rightarrow U(1)^r$  %(naive) %Abelian monopole

\item
   \textbf{field decomposition} (by Cho\cite{Cho80}, Duan and Ge \cite{DG79}, Faddeev and Niemi \cite{FN99}, Shabanov \cite{Shabanov99}, Kondo, Murakami and Shinohara \cite{KMS05,KMS06,Kondo06}): 

gauge-invariant decomposition of the gluon field for separating the dominant mode for confinement %(involved)  
%[spin-charge separation]
\end{enumerate} 
It is  very important to answer the question of how to define the gauge-invariant chromomagnetic monopole   in the Yang-Mills gauge theory without scalar fields,  which should be discriminated from the 't Hooft-Polyakov magnetic monopole in the Georgi-Glashow model. 
However, the details of this issue will be discussed elsewhere, since it is not the main issue to be discussed in this paper.

For the dual superconductor picture for the Yang-Mills theory vacuum to be true,  the chromomagnetic monopole condensation must  give a more  stable vacuum than the perturbative one. 
In view of this, Savvidy \cite{Savvidy77} has argued based on the general analysis of the renormalization group equation that the  \textbf{dynamical generation of chromomagnetic field} should occur in the Yang-Mills theory, i.e., a non-Abelian gauge theory with asymptotic freedom.  Indeed, Savvidy has shown that the vacuum with a non-vanishing  \textbf{homogeneous chromomagnetic field} strength, the so-called  \textbf{Savvidy vacuum}, has lower energy density than the perturbative vacuum with zero  chromomagnetic field.  
The one-loop effective potential $V(H)$ of the homogeneous chromomagnetic field $H$  obtained in \cite{Savvidy77} for  $SU(2)$ Yang-Mills theory is 
\begin{equation}
V_{\rm Savvidy}(H)
 =\frac12 H^2
 %\left[ 1
 -\frac{\beta_0 g^2}{16\pi^2}
 \frac12 H^2  \left(\ln\frac{gH}{\mu^2} + c \right)
% \right]
, \
%\beta_0 :=-\frac{22}3 <0 
\end{equation}
with
$
\beta_0 :=-\frac{22}3 <0 ,
$
and a constant $c$.
Then the effective potential $V(H)$ has an absolute minimum at $H=H_0 \ne 0$ away from $H=0$.
%The {chromomagnetic condensation} gives more stable vacuum than the perturbative one. 
%Since condensates are scalar, it seems like a good first approximation that the vacuum contains some non-zero but homogeneous field which gives rise to these condensates. 
%\footnote{
%  G.K. Savvidy,
%Infrared instability of the vacuum state of gauge theories and asymptotic freedom,
%Phys. Lett. B {\bf 71}, 133--134 (1977).
%} 

Immediately after his proposal, however, N.K.Nielsen and Olesen \cite{NO78} have pointed out  that  the effective potential $V(H)$ of the homogeneous chromomagnetic field $H$, when calculated carefully at one-loop level in the perturbation theory under the background gauge, develops  an imaginary part in the one-loop effective potential: 
\begin{equation}
%\hspace*{-100mm}
V_{\rm NO}(H)
 =\frac12 H^2
 %\left[1 
 -\frac{\beta_0 g^2}{16\pi^2} \frac12 H^2
   \left(\ln\frac{gH}{\mu^2} + c  \right)
% \right]
+   i \ \frac{g^2 H^2}{8\pi}    ,
\end{equation}
in addition to the real part which agrees exactly with  the prediction of the renormalization group equation, i.e., the Savvidy's result. 
This is called the \textbf{Nielsen-Olesen (NO) instability} of the Savvidy vacuum.
 The presence of the pure imaginary part implies that the Savvidy vacuum gets unstable due to gluon--antigluon pair annihilation. 
%It should be remarked that both effective potentials satisfy the renormalization group equation and the imaginary part cannot be excluded using the renormalization group equation. 
%\footnote{
%  N.K. Nielsen and P. Olesen,
%An unstable Yang-Mills field mode,
%Nucl. Phys. B {\bf 144}, 376--396 (1978).
%} 

This result is easily understood based on the following observation. 
In the homogeneous external chromomagnetic field $H$,  the energy eigenvalue $E_n$ of the massless ({\it off-diagonal}) gluons with the spin $S=1$  ($S_z=\pm 1$)   is given by
\begin{equation}
 E_n^\pm=\sqrt{p_\perp^2+2gH (n+1/2) + 2gH   S_z}  \ (n=0,1,2,\cdots) ,
\end{equation}
where $p_{\perp}$ denotes the momentum in those space-time directions that are not affected by the magnetic field  and the index $n$ is a discrete quantum number that labels the \textbf{Landau levels}.
Then the NO instability is  understood as originating from the \textbf{tachyon mode} with $n=0$ and  $S_z=-1$ (or the lowest Landau level for the gluon with  spin one antiparallel to the external chromomagnetic field), 
since 
\begin{equation}
E_0^-=\sqrt{p_\perp^2-gH} ,
\end{equation}
becomes pure imaginary when $p_\perp^2<gH$.
In other words, the NO instability of the Savvidy vacuum with homogeneous chromomagnetic condensation is due to the existence of the  tachyon mode  corresponding to the lowest Landau level which is realized  in the homogeneous chromomagnetic field.

It is instructive to compare the Yang-Mills theory with QED to understand the NO instability correctly, since the instability of QED (an Abelian gauge theory without asymptotic freedom) under the applied electric field is well known. 
%On the other hand, 
In QED, the non-zero magnetic field does not lower the vacuum energy and hence no magnetic condensation occurs, while the electric field causes electron--positron pair creation, destabilizing the QED vacuum. 
The chromoelectric field destabilizes the vacuum also in Yang-Mills theory. 
Therefore, the instability of the Yang-Mills vacuum under the chromomagnetic field is quite different from the instability of QED.

A way to circumvent the NO instability  is to introduce the \textbf{magnetic domains} (domain structure) with a finite extension into the vacuum \cite{Copenhagen}. 
The physical vacuum in Yang-Mills theory is split into an infinite number of domains with macroscopic extensions. 
Inside each such domain there is a nontrivial configuration of the chromomagnetic field and the tachyon mode does not appear in the domain supporting  $p_\perp^2>gH$. 
This resolution for  the NO  instability of Yang-Mills theory  is called the \textbf{Copenhagen vacuum} or \textbf{Spaghetti vacuum}. 
%It is instructive to recall that the vortices in a type II superconductor are neatly arranged into a hexagonal or occasionally square lattice  \cite{superconductor}.
%The Copenhagen vacuum breaks the Lorentz invariance and color invariance explicitly.
%Averaging over all domains results in a zero background chromomagnetic field, hence color and Lorentz symmetries are not broken.%

%The NO instability  was derived based on the one-loop calculation of the effective potential.  Therefore,  some people consider it as indicating unreliability of the (lowest-order) loop calculation, i.e., artifact of the approximation.
% \cite{Kennaway04}.  
%However, no one has demonstrated that the inclusion of higher order terms cures the instability.% 
%\footnote{
%Moreover, the similar problem exists also in the supersymmetric Yang-Mills theory in which the higher-order loop corrections are absent, since the covariantly constant background field strength is not supersymmetric.
% \cite{Kay83}. 
%}
%Nevertheless, it is not yet full investigated so far. 
%Can the instability be resolved in the one-loop level by a new mechanism? 

What type of vacuum is allowed and preferred in the Yang-Mills theory is an important question related to the physical picture of quark confinement.  
We can say that the NO instability is an infrared problem in the non-Abelian gauge theory. 
%The Copenhagen vacuum is a well-done model of the Yang-Mills vacuum.  
The domain structure introduces an infrared cutoff which prevents  the momenta from taking the smaller values causing the instability. 
However, it is quite complicated to work out the dynamics of the Yang-Mills theory on the concrete inhomogeneous background. 
Therefore, there have been a lot of works trying to overcome the NO instability for the homogeneous chromomagnetic field
\cite{Kay83,KPV03,YC80,Leutwyler80,Adler81,Schanbacher82,Flory83,DR83,Dunne04,Kennaway04,KKP05,Parthasarathy10,CP01,Cho03,CWP04,Cho13,Kondo04,Kondo05,KKMSS05,VV08}.

In view of these, we reexamine the NO instability in the $SU(2)$ Yang-Mills theory
\footnote{
In order to study the true QCD vacuum, we must discuss the $SU(3)$ gauge group. 
However,  $SU(3)$ case is more difficult from a technical viewpoint than the $SU(2)$ case. In this paper, therefore, we discuss the $SU(2)$ toy model and postpone the physical $SU(3)$  case  in a subsequent paper. 
It should be remarked that the physically interesting case of the $SU(3)$ gauge group cannot be obtained by a simple group-theoretical extension of the $SU(2)$ case and that the different results could be obtained in the case of $SU(3)$, which is suggested from a formal consideration \cite{KSM08}.  
}
  in the framework of the \textbf{functional renormalization group} (FRG)  \cite{Wetterich93,FRG} as a realization of the \textbf{Wilsonian renormalization group} \cite{WRG}.
The FRG enables us to examine the effects caused by changing the infrared cutoff in a systematic way.

In this paper we follow the methods developed for FRG in \cite{RW94,RW97,Gies02,EGP11}.
We point out the following results. 

\begin{enumerate}
\item 
  The \textbf{Nielsen-Olesen instability} in the effective potential $V(H)$ for the homogeneous chromomagnetic field $H$, i.e., the imaginary part Im $V(H)$ of $V(H)$   disappears (or is absent from the beginning) in the framework of the  FRG. 
  (Therefore, the Nielsen-Olesen instability is an artifact   of the one-loop calculation in the perturbation theory and it disappears in the non-perturbative framework beyond the perturbation theory.)

\item 
 However, this result does not necessarily guarantee the automatic existence of the non-trivial homogeneous chromomagnetic field $H_0 \neq 0$ as the minimum of the effective potential $V(H)$, such that 
\begin{equation}
V(H_0) < V(H=0)=0.
\end{equation}
(Therefore, the absence of the Nielsen-Olesen instability and the existence of the non-trivial minimum for the homogeneous chromomagnetic field in the effective potential are different problems to be considered independently.)
\footnote{
This question was studied in \cite{EGP11} giving the answer in the affirmative by using the self-dual background, which does not suffer from the instability from the beginning. 
This work is quite interesting, but does not answer other questions raised here for other choices of the background. 
See Conclusion and Discussion. 
}

\item 
 As a physical mechanism for maintaining the stability even for the small infrared cutoff, we propose the \textbf{dynamical mass generation} for the off-diagonal gluons (and off-diagonal   ghosts), which is related to the Becchi-Rouet-Stora-Tyutin (BRST)-invariant \textbf{vacuum condensation of mass-dimension two}  \cite{GSZ01,Kondo01,Kondo03,KMSI02}.  
This gives a consistent picture compatible with the absence of the instability. 
(This leads to the \textbf{Abelian dominance} \cite{tHooft81,EI82}: in the string tension extracted from the Wilson loop  average \cite{SY90} and exponential falloff of the off-diagonal gluon propagators \cite{AS99,BCGMP03,MCM06,CDG-numerical} as well as the magnetic monopole dominance \cite{SNW94} in the Maximal Abelian gauge \cite{KLSW87}.)
\end{enumerate}

This paper is organized as follows.
In Sec. II, we consider the complex-valued flow equation for the effective average action and decompose it into the real and imaginary parts. 
We show that the flow equation has a solution with vanishing  imaginary part of the  effective average action for any value of the infrared cutoff $\Lambda$, corresponding to the fixed point. 

In Sect. III, we derive explicitly the flow equation for the effective average action  in the chromomagnetic background. 
As an infrared regulator, we use the mass type infrared cutoff function to give a closed form for the flow equation. 
By removing the ultraviolet divergence due to this choice of the infrared function, we obtain a flow equation with the infrared cutoff that is free from the ultraviolet divergence. 

In Sect. IV, we show the absence of the NO instability, i.e., vanishing of the imaginary part of the effective average action. 
This is done based on an approximate solution obtained by solving the flow equation for   large values of the infrared cutoff $\Lambda$. 

In Sect. V, we discuss the mass generation for the off-diagonal gluons and ghosts due to the vacuum condensation of mass-dimension two, which is BRST invariant. 
Moreover, we argue the relationship between   stability and mass generation. 

The final section is devoted to   conclusions and discussions.

%%%%%%%%%%%%%%%%%%%%%%%%%%%%%%%%%%%%%%%%%%%%%%%%%

\section{Complex-valued flow equation }

%%%%%%%%%%%%%%%%%%%%%%%%%%%%%%%%%%%%%%%%%%%%%%%%%

The \textbf{effective average action} $\Gamma_\Lambda$ with the infrared cutoff $\Lambda$ is obtained by solving the flow equation \cite{Wetterich93}: %(in a symbolic notation) 
\begin{equation}
\partial_{t} \Gamma_{\Lambda} = \frac{1}{2} {\rm STr} \bigl[ ( \Gamma_{\Lambda}^{(2)} + R_{\Lambda} )^{-1} \cdot \partial_{t} R_{\Lambda} \bigr] , \ 
\partial_{t} := \Lambda \frac{d}{d \Lambda} ,
\end{equation}
where ${\rm STr}$ denotes the ``supertrace'' introduced for   writing both  commuting fields (e.g., gluons) and anticommuting fields (e.g., quarks and the Faddeev-Popov ghosts), $R_\Lambda^\Phi$ is the \textbf{infrared cutoff function} for the field $\Phi$  introduced as the \textbf{infrared regulator term}  in the form:
\begin{equation}
\int \Phi^\dagger R_\Lambda^\Phi \Phi ,
%for which the infrared regulators $R_\Lambda^\Phi$ are introduced, 
\end{equation}
and $\Gamma_\Lambda^{(2)}$ denotes the second  derivatives  of $\Gamma_{\Lambda}$ with respect to the field variables $\Phi$, 
\begin{equation}
(\Gamma_\Lambda^{(2)})_{\Phi^\dagger \Phi} = \frac{\overrightarrow{\delta}}{\delta \Phi^\dagger} \Gamma_\Lambda \frac{\overleftarrow{\delta}}{\delta \Phi} ,
\end{equation}
corresponding to the inverse exact propagator at the scale $\Lambda$. 
The ordinary \textbf{effective action} $\Gamma$ as the generating functional of the one-particle irreducible vertex functions is obtained in the limit $\Lambda  \downarrow 0$: $\Gamma = \lim_{\Lambda \downarrow 0} \Gamma_\Lambda$.

We consider the complex-valued effective average action $\Gamma_{\Lambda} = \Gamma_{\Lambda}^{\rm R} + i \Gamma_{\Lambda}^{\rm I}$ which is decomposed into the real part $\Gamma_{\Lambda}^{\rm R}:=\text{Re} \Gamma_\Lambda$ and the imaginary part $\Gamma_{\Lambda}^{\rm I}:=\text{Im} \Gamma_\Lambda$.
Then it is shown (see Appendix A) that the flow equation is decomposed into two parts:
\begin{align}
 & \partial_{t} \Gamma_{\Lambda}^{\rm R} 
\nonumber\\
=& \frac{1}{2} {\rm STr} \Big\{ \bigl[ ( \Gamma_{\Lambda}^{\rm R (2)} + R_{\Lambda} )^{2} + ( \Gamma_{\Lambda}^{\rm I (2)})^{2} \bigr]^{-1} 
%\nonumber\\& \times
 ( \Gamma_{\Lambda}^{\rm R (2)} + R_{\Lambda} ) \partial_{t} R_{\Lambda}   \Big\} , 
\\
 & \partial_{t} \Gamma_{\Lambda}^{\rm I} 
\nonumber\\
=&  - \frac{1}{2} {\rm STr} \left\{ \bigl[ ( \Gamma_{\Lambda}^{\rm R (2)} + R_{\Lambda} )^{2} + ( \Gamma_{\Lambda}^{\rm I (2)})^{2} \bigr]^{-1} \Gamma_{\Lambda}^{\rm I (2)} \partial_{t} R_{\Lambda} \right\} .
\end{align}

%-------- Figure ----------------------------
\begin{figure}[t]
\begin{center}
\includegraphics[scale=0.35]{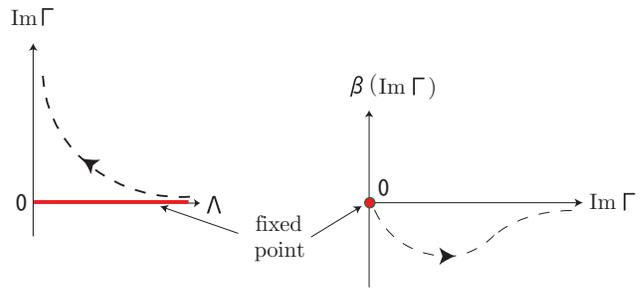}
\quad\quad
\end{center}
\vskip -0.5cm
\caption{ 
(Left panel) Imaginary part $\text{Im} \Gamma_\Lambda$ of the average effective action as a function of $\Lambda$,
(Right panel)   $\beta$ function as a function of $\text{Im} \Gamma_\Lambda$.
Here $\text{Im} \Gamma_\Lambda=0$ corresponds to a fixed point of the renormalization group. 
}
\label{fig:complex-action} 
\end{figure}
%---------------------------------------------

We find that the flow equation has a remarkable property: \textit{the identically vanishing imaginary part $\Gamma_{\Lambda}^{\rm I}:=\text{Im} \Gamma_\Lambda  \equiv 0 $ is an exact solution corresponding to a ``\textbf{fixed point}''}:
\begin{equation}
\text{Im} \Gamma_\Lambda \equiv 0  \ \ \text{for any value of} \ \ \Lambda ,
\end{equation}
in sharp contrast with the real part.
See Fig.~\ref{fig:complex-action} for the behavior of the ``beta''-function of  $\Gamma_{\Lambda}^{\rm I}$ defined by 
\begin{equation}
\beta(\Gamma_{\Lambda}^{\rm I}) :=\partial_{t} \Gamma_{\Lambda}^{\rm I} .
\end{equation}
If $\Gamma_{\Lambda}^{\rm I} \not= 0$ for a certain value of $\Lambda$, it does not maintain the same value, i.e., $\beta(\Gamma_{\Lambda}^{\rm I})  \not =0$.
\footnote{
If the right-hand side of the flow equation for  $\Gamma_{\Lambda}^{\rm I}$ was linear in $\Gamma_{\Lambda}^{\rm I}$,  a stronger statement would be derived: 
If $\Gamma_{\Lambda}^{\rm I}$ vanishes for some particular value of $\Lambda$, e.g., for $\Lambda = \Lambda_{0}$, then $\Gamma_{\Lambda}^{\rm I}$ vanishes automatically for any other value of $\Lambda \le \Lambda_{0}$, in particular for $\Lambda \to 0$, provided that $\Gamma_{\Lambda}$ and $\Gamma_{\Lambda_{0}}$ are related by integrating (or solving) the flow equation:
\begin{equation}
 \Gamma_{\Lambda}^{\rm I} = 0 \ \text{at} \ \Lambda = \Lambda_{0} \Longrightarrow \Gamma_{\Lambda}^{\rm I} = 0 \ \text{for} \ \Lambda < \Lambda_{0} .
\end{equation}
}

Thus the problem of showing the absence of the imaginary part $\text{Im}\Gamma =\lim_{\Lambda \downarrow 0} \text{Im}\Gamma_\Lambda$ in the effective action $ \Gamma=\lim_{\Lambda \downarrow 0} \Gamma_\Lambda$ is reduced to proving the vanishing of the imaginary part  Im$ \Gamma_\Lambda  $ in the effective average action $ \Gamma_\Lambda $ for a sufficiently large value of $\Lambda$:
\begin{equation}
\text{Im} \Gamma_\Lambda  =0 \ \ \text{for a certain value of} \ \ \Lambda \gg 1.
\end{equation}
When $\Gamma_{\Lambda}^{\rm I} = 0$, the flow equation for $\Gamma_{\Lambda}^{\rm R}$ turns into the standard flow equation.

%%%%%%%%%%%%%%%%%%%%%%%%%%%%%%%%%%%%%%%%%%%%%%%%%

\section{Flow equation in the chromomagnetic background}

%%%%%%%%%%%%%%%%%%%%%%%%%%%%%%%%%%%%%%%%%%%%%%%%%

We consider the $D$-dimensional Euclidean Yang-Mills theory. 
We decompose the $SU(2)$ Yang-Mills field $\mathscr{A}_\mu=\mathscr{A}_\mu^A T^A$ into the \textbf{background field} $\mathscr{V}_\mu=\mathscr{V}_\mu^A T^A$ and \textbf{the quantum fluctuation field}  $\mathscr{X}_\mu=\mathscr{X}_\mu^A T^A$ where $T^A=\frac{1}{2} \sigma^A$ with $\sigma^A$ being the Pauli matrices ($A=1, 2, 3$):
\begin{equation}
\mathscr{A}_\mu^A =\mathscr{V}_\mu^A + \mathscr{X}_\mu^A  \quad   (A=1, 2, 3)  .
\end{equation}
%where $\mathscr{A}_\mu=\mathscr{A}_\mu^A T^A$, $\mathscr{V}_\mu=\mathscr{V}_\mu^A T^A$, $\mathscr{X}= \mathscr{X}_\mu^A T^A$ and $T^A=\frac{1}{2} \sigma^A$ with $\sigma^A$ being the Pauli matrices ($A=1, 2, 3$).

We can choose without loss of generality \cite{KMS05}  the diagonal field $V_\mu$ as the background field:
\begin{equation}
\mathscr{V}^A_\mu (x) = \delta^{A3} V_\mu (x) ,
\label{V}
\end{equation}
and the off-diagonal field $A_\mu^a$ ($a=1, 2$)  as the quantum fluctuation field:
\begin{equation}
\mathscr{X}_\mu^A (x) =\delta^{Aa} A_\mu^a (x) \ , \ \ (a=1, 2) ,
\end{equation}
which means that 
\begin{align}
\mathscr{A}_\mu =& V_\mu \frac{\sigma^3}{2}+ A_\mu^a \frac{\sigma^a}{2}  %\ (a=1,2)  %\quad \text{or}  \quad
%\nonumber\\
\ \text{or} \
\mathscr{A}_\mu^3 =  V_\mu , \ \mathscr{A}_\mu^a=A_\mu^a  \ (a=1,2) . 
\end{align}

For the diagonal gauge field $V$ and the off-diagonal gauge field $A$, the   Yang-Mills Lagrangian has the interaction terms of the type:
$VAA$, $VVAA$ and $AAAA$, 
while the gauge-fixing (GF) and the associated Faddeev-Popov (FP) term for the maximal Abelian (MA) gauge (defined shortly) has the interactions of the type:
$VAA$ and $VVAA$.
The effective potential $\Gamma(V)$ of   $V$ is obtained from  the diagrams with the external legs of $V$  by integrating all the  internal lines that are connected through the possible interaction vertices.
For the one-loop effective potential, accordingly, it is easy to see that only the internal lines of  $A$ are allowed, which implies that there is no fluctuating diagonal field $A$ to be integrated out. 
For large $\Lambda$, the deviation from the one-loop is not so significant and the fluctuating diagonal field can be neglected.% 
\footnote{
Such a distinction between the diagonal and off-diagonal fields can be partially justified and has been used so far based on the results of   numerical simulations of $SU(2)$ Yang-Mills theory on a lattice.  
Beyond one-loop, of course, this simplification is not allowed and we must integrate out the diagonal fluctuation field.
In fact,  in order to show the confinement/deconfinement transition at finite temperature, the diagonal fluctuations play the most important and essential role, as first shown in   \cite{MP08}
%\\
%\bibitem{MP08}
%F. Marhauser and J.M. Pawlowski,
%Confinement in Polyakov gauge,
%arXiv:0812.1144[hep-ph].
%\\
%which was added in the reference in the revised version, 
and  confirmed  in   \cite{Kondo10}
%Kondo (2010) 
using the same framework as that of this paper. 
Such a contribution will be included in the subsequent work where the interplay between the existence/non-existence of chromomagnetic condensation and confinement/deconfinement at finite temperature will be investigated. 
In view of these, this paper is the first attempt  towards the thorough analysis of the stability of the chromomagnetic condensation in the QCD vacuum. 
}

In what follows, we prepare the diagonal field $V_\mu (x)$ of the form:
\begin{equation}
 V_\mu(x) = \frac{1}{2} x_\nu H_{\nu \mu}  ,
%\quad H_{\alpha \beta} = \partial_\alpha V_\beta (x) -\partial_\beta V_\alpha (x)
\end{equation}
so that   the  $x$-independent \textbf{homogeneous  background field strength} is realized:
\begin{align}
\mathscr{F}_{\mu \nu}^A [\mathscr{V}](x) 
:=& \partial_\mu \mathscr{V}_\nu^A(x) -\partial_\nu \mathscr{V}_\mu^A(x) +\epsilon^{ABC} \mathscr{V}_\mu^B(x) \mathscr{V}_\nu^C(x)
 \nonumber\\
%\mathscr{F}_{\mu \nu}^A [\mathscr{V}](x) 
%:= \partial_\mu \mathscr{V}_\nu^A(x) -\partial_\nu \mathscr{V}_\mu^A(x) +\epsilon^{ABC} \mathscr{V}_\mu^B(x) \mathscr{V}_\nu^C(x)
=& \delta^{A3} \left( \partial_\mu V_\nu(x) -\partial_\nu V_\mu(x) \right) =\delta^{A3} H_{\mu \nu}.
\end{align}
Then the background field strength  realizes the (homogeneous) chromomagnetic field: 
\begin{equation}
 \bm{H}=(H_1,H_2,H_3)
\end{equation}
by choosing the non-vanishing components as
\begin{align}
H_{23} =& - H_{32} := H_{1} , \ H_{31} = - H_{13} := H_{2} , \ 
\nonumber\\
H_{12} =& - H_{21} := H_{3}  .
\end{align}

The \textbf{total  effective average action}  $\Gamma_\Lambda$  is specified by giving the gauge-invariant part $\Gamma_\Lambda^{\rm inv}$, the GF part $\Gamma_\Lambda^{\rm GF}$ and the associated  FP ghost part $\Gamma_\Lambda^{\rm FP}$:
\begin{equation}
\Gamma_\Lambda=\Gamma_\Lambda^{\rm inv} +\Gamma_\Lambda^{\rm GF} + \Gamma_\Lambda^{\rm FP}.
\end{equation}

We choose the \textbf{background gauge} \cite{Abott81} as the gauge fixing condition to maintain the gauge invariance for the background field. 
In the above choice for the background field (\ref{V}), the background gauge reduces to the \textbf{maximal Abelian} (MA) gauge: 
\begin{equation}
 F^{a} := \mathscr{D}_\mu^{ab} [V] A_\mu^b =0 , 
%\end{equation}
%where
%\begin{equation}
\quad
\mathscr{D}_\mu^{ab}[V]  := \partial_\mu \delta^{ab}-g \epsilon^{ab3} V_\mu .
\end{equation}
Then  \textbf{the gauge-fixing term} is given by
\begin{equation}
\Gamma^{\rm GF} =\int d^Dx \frac{1}{2\alpha} \left( \mathscr{D}_{\mu}^{ab} [V] A_\mu^b \right)^2 , 
\end{equation}
where $\alpha$ denotes the gauge-fixing parameter. 
This $\Gamma_{\rm GF}$ is obtained by integrating out the Nakanishi-Lautrup field $N^a$ from
\begin{equation}
\Gamma^{\rm GF} = \int d^Dx  \left\{   N^a  \left( \mathscr{D}_\mu^{ab} [V] A_\mu^b \right)+\frac{\alpha}{2}  N^a N^a  \right\} .
\end{equation}

The \textbf{FP ghost term} is determined according to the standard procedure (see e.g., \cite{Kondo97}) as 
\begin{align}
\Gamma^{\rm FP} =   \int d^Dx  \{ & 
  i \bar{C}^{a} \mathscr{D}_{\mu}^{ab} [V] \mathscr{D}_{\mu}^{bc} [V] C^{c} 
\nonumber\\ & 
- g^{2} \epsilon^{ab3} \epsilon^{cd3}  i\bar{C}^{a} C^{d} A_{\mu}^{b} A_{\mu}^{c}  
\nonumber\\ &
+ i \bar{C}^{a} g\epsilon^{ab3} (\mathscr{D}_{\mu}^{bc} [V] A_{\mu}^{c}) C^{3} \} .
\end{align}

For the gauge-invariant part $\Gamma_\Lambda^{\rm inv}$, we adopt the  ansatz,   a function $W_\Lambda$  of the gauge-invariant term $\Theta$ constructed from the field strength $ \mathscr{F}_{\mu \nu}^A [\mathscr{A}]:= \partial_\mu \mathscr{A}_\nu^A-\partial_\nu \mathscr{A}_\mu^A+\epsilon^{ABC} \mathscr{A}_\mu^B \mathscr{A}_\nu^C $:
 \begin{equation}
\Gamma_\Lambda^{\rm inv} =\int d^Dx W_\Lambda \left( \Theta (x) \right) , 
\quad
\Theta:= \frac{1}{4} \left( \mathscr{F}_{\mu \nu}^A [\mathscr{A}] \right)^2 .
\end{equation}
 $\Theta$ is decomposed as  \cite{Kondo10}
\begin{align}
\Theta =& \frac{1}{4} \left( \mathscr{F}_{\mu \nu}^A [\mathscr{V}] \right)^2 +
 \frac{1}{2} A^{\mu a} \left( Q_{\mu \nu}^{ab}+\mathscr{D}_\mu^{ac}[V] \mathscr{D}_\nu^{cb} [V] \right) A^{\nu b} 
%\notag \\&
\nonumber\\ &+\frac{1}{4} \left( \epsilon^{3ab} A_\mu^a A_\nu^b \right)^2 ,
\end{align}
where
\begin{align}
  Q_{\mu \nu}^{ab} :=& - \left(\mathscr{D}^2 \right)^{ab} \delta_{\mu \nu} +2g \epsilon^{ab} H_{\mu \nu} ,
\nonumber\\  
\left(\mathscr{D}^2 \right)^{ab} :=& \mathscr{D}_\rho^{ac} [V] \mathscr{D}_\rho^{cb} [V] .
\end{align}
In the vanishing off-diagonal field limit $A_\mu^a \to0$, $\Theta$ is reduced  to
\begin{align}
\Theta|_{A=0}=& \frac{1}{4} \left( \mathscr{F}_{\mu \nu}^A [\mathscr{V}](x) \right)^2 
=\frac{1}{4} \left( \partial_\mu V_\nu(x)  - \partial_\nu V_\mu(x)  \right)^2
\nonumber\\   
=& \frac{1}{2} H^2 ,
\\
 H :=&   \sqrt{\bm{H}^2} = \sqrt{\frac{1}{2} H_{\alpha \beta} H_{\alpha \beta}}>0  .
\end{align}
%where 
%\begin{equation}
%\end{equation}

The off-diagonal gluon fields $A_\mu^a$ (and  off-diagonal ghost fields $C^a$, $\Bar{C}^a$) should be integrated out in the framework of the FRG following the idea of  the Wilsonian renormalization group.
%\footnote{
%\bibitem{MP08}
%F. Marhauser and J.M. Pawlowski,
%Confinement in Polyakov gauge,
%arXiv:0812.1144[hep-ph].
%}
For this purpose, we  introduce the \textbf{infrared regulator term} $\Delta S_\Lambda$ for the off-diagonal gluon $A_\mu^a$ and off-diagonal ghosts $C^a$, $\Bar{C}^a$ by
\begin{align}
\Delta S_\Lambda =& \int_p \Big[ \frac{1}{2} A_\mu^a (p) R_{\Lambda, \mu \nu} (p^2) \delta^{ab} A_\nu^b (p) 
\nonumber\\ &
+\Bar{C}^a (p) R_\Lambda (p^2) \delta^{ab} C^b (-p) \Big] \ (a,b=1,2),
\end{align}
where 
\begin{equation}
 \int_p := \int \frac{d^Dp}{(2\pi)^D}
\end{equation}
  denotes the integration over the $D$-dimensional momentum space.
We choose  the infrared cutoff function with the structure:
\begin{equation}
R_{\Lambda, \mu \nu} (p^2) = \delta_{\mu \nu}  R_\Lambda (p^2) .
\end{equation}

We adopt the \textbf{proper-time form} of the flow equation \cite{LP02}:
\begin{equation}
\partial_t \Gamma_\Lambda =\int^\infty_0 d\tau \frac{1}{2} {\rm STr} \left[ e^{-\tau \left(\Gamma_\Lambda^{(2)}+R_\Lambda \right)} \partial_t R_\Lambda \right].
\end{equation}
After performing the mode decomposition according to the projection method \cite{RW94,RW97, Gies02,EGP11},   the flow equation reads
\begin{align}
\partial_{t} \Gamma_{\Lambda} =& \frac{1}{2} \int_{0}^{\infty} d \tau \ \Omega^{-1} {\rm Tr} \bigl[ e^{- \tau ( W_{\Lambda}^{\prime} Q + R_{\Lambda}^{\rm gluon} )} \cdot \partial_{t} R_{\Lambda}^{\rm gluon} \bigr]  
%(\text{transverse gluons})
\nonumber\\&
- \frac{1}{2} \int_{0}^{\infty} d \tau \ \Omega^{-1} {\rm Tr} \bigl[ e^{- \tau (  - W_{\Lambda}^{\prime} \mathscr{D}^2 + R_{\Lambda}^{\rm gluon} )} \cdot \partial_{t} R_{\Lambda}^{\rm gluon} \bigr]  
%(\text{scalar gluon})
\nonumber\\&
+ \frac{1}{2} \int_{0}^{\infty} d \tau \ \Omega^{-1} {\rm Tr} \bigl[ e^{- \tau (  -\alpha_\Lambda^{-1}\mathscr{D}^2 + R_{\Lambda}^{\rm gluon} )} \cdot \partial_{t} R_{\Lambda}^{\rm gluon} \bigr]   
%(\text{longitudinal gluon})
\nonumber\\&
- \int_{0}^{\infty} d \tau \ \Omega^{-1} {\rm Tr} \bigl[ e^{- \tau ( - \tilde{Z}_{\Lambda} \mathscr{D}^2 + R_{\Lambda}^{\rm ghost} )} \cdot \partial_{t} R_{\Lambda}^{\rm ghost} \bigr]   
%(\text{ghosts })
 ,
\end{align}
where we have defined
$
 W'_\Lambda (\Theta) := \frac{d}{d\Theta} W_\Lambda (\Theta) .
$
Here we have introduced  the \textbf{wavefunction renormalization constants}:
\begin{equation}
Z_\Lambda =Z_\Lambda^{\rm gluon}  , \ 
\Tilde{Z}_\Lambda=Z_\Lambda^{\rm ghost} .
\end{equation}
In this derivation, we have adopted the truncation --- neglecting the four-point interactions 
among the off-diagonal gluons and off-diagonal ghosts 
$
- g^{2} \epsilon^{ab3} \epsilon^{cd3}  i\bar{C}^{a} C^{d} A_{\mu}^{b} A_{\mu}^{c}  
$, 
which do not couple to the background field $V_\mu$.

The spectrum sum is obtained from eigenvalues of the respective operator. 
The \textbf{covariant Laplacian} $- \left( \mathscr{D}_{\rho} [\mathscr{V}] \right)^{2}$ with the background field $\mathscr{V}$ which gives the (covariant constant) uniform chromomagnetic field $H$ has the spectrum:
\begin{equation}
{\rm Spect} \bigl[ - \mathscr{D}_{\rho}^{2} [\mathscr{V}] \bigr] = p_{\perp}^{2} + ( 2 n + 1 ) gH , \ ( n = 0 , 1 , \cdots ),
\end{equation}
where $p_{\perp}$ denotes the $(D-2)$ dimensional (Fourier) momentum in those space-time directions that are not affected by the magnetic field (say, orthogonal to $1-2$ plane) and the index $n$ is a discrete quantum number that labels the Landau levels.
We take into account the fact that the density of states is $\frac{gH}{2\pi}$ for the \textbf{Landau levels}.

Moreover, the operator $Q_{\mu\nu}^{ab}$ with the same background field $\mathscr{V}$ has the spectrum:
\begin{align}
{\rm Spect} \bigl[ Q_{\mu\nu}^{ab} \bigr] =&
\left\{ \begin{array}{c}
p_{\perp}^{2} + ( 2 n + 1 ) gH \\
p_{\perp}^{2} + ( 2 n + 3 ) gH \\
p_{\perp}^{2} + ( 2 n - 1 ) gH
\end{array} \right.   {\rm multiplicity} \ \left. \begin{array}{c}
(D - 2) \\
1 \\
1 
\end{array} \right. 
\nonumber\\ &
( n = 0 , 1 , \cdots ),
\end{align}
where the last term contains the \textbf{Nielsen-Olesen unstable mode}  for $n=0$, i.e., 
\begin{equation}
 p_{\perp}^{2}-gH , 
\end{equation}
which becomes a \textbf{tachyonic mode} for small momenta $p_{\perp}^{2} < gH$.

The respective trace without the infrared regulator $R_\Lambda$ is easily obtained \cite{RW94,RW97,Gies02}:
\begin{align}
&  \Omega^{-1} {\rm Tr} [ e^{- \tau ( W_{\Lambda}^{\prime} Q ) } ]
\nonumber\\
=& \frac{NgH}{(4\pi)^{\frac{D}{2}}} ( \tau W_{\Lambda}^{\prime} )^{1 - \frac{D}{2}}  \biggl[ \frac{D}{\sinh (\tau W_{\Lambda}^{\prime} gH)} + 4 \sinh ( \tau W_{\Lambda}^{\prime} gH ) \biggr]
\nonumber\\
=& \frac{NgH}{(4\pi)^{\frac{D}{2}}} ( \tau W_{\Lambda}^{\prime} )^{1 - \frac{D}{2}}    
\nonumber\\ & \times
 \frac{ 2(D-2)  e^{-\tau W_{\Lambda}^{\prime} gH} + 2 e^{-3\tau W_{\Lambda}^{\prime} gH} +  2 e^{\tau W_{\Lambda}^{\prime} gH} }{1-e^{-2\tau W_{\Lambda}^{\prime} gH}}  
 ,
%\end{align}
%and
%\begin{align}
\nonumber\\
 & \Omega^{-1} {\rm Tr} [ e^{- \tau ( -W_{\Lambda}^{\prime} \mathscr{D}^2 ) } ]
%\nonumber\\
=  \frac{NgH}{(4\pi)^{\frac{D}{2}}} ( \tau W_{\Lambda}^{\prime} )^{1 - \frac{D}{2}}  \biggl[ \frac{1}{\sinh (\tau W_{\Lambda}^{\prime} gH)}  \biggr] 
\nonumber\\
=& \frac{NgH}{(4\pi)^{\frac{D}{2}}} ( \tau W_{\Lambda}^{\prime} )^{1 - \frac{D}{2}}  \biggl[ \frac{2 e^{-\tau W_{\Lambda}^{\prime} gH} }{ 1-e^{-2\tau W_{\Lambda}^{\prime} gH} }  \biggr] ,
\nonumber\\
 &  \Omega^{-1} {\rm Tr} [ e^{- \tau ( -\tilde{Z}_{\Lambda}  \mathscr{D}^2 ) } ]
%\nonumber\\
=  \frac{NgH}{(4\pi)^{\frac{D}{2}}} ( \tau \tilde{Z}_{\Lambda} )^{1 - \frac{D}{2}}  \biggl[ \frac{1}{\sinh (\tau \tilde{Z}_{\Lambda}  gH)}  \biggr] 
\nonumber\\
=& \frac{NgH}{(4\pi)^{\frac{D}{2}}} ( \tau \tilde{Z}_{\Lambda} )^{1 - \frac{D}{2}}  \biggl[ \frac{ 2 e^{-\tau \tilde{Z}_{\Lambda} gH} }{1-e^{-2\tau \tilde{Z}_{\Lambda} gH} }  \biggr] ,
\end{align}
where $N=2$ for $SU(2)$.

In order to  obtain the closed analytical form for the solution and to compare the FRG calculations with the loop calculations, 
we choose the momentum-independent \textbf{infrared regular of the mass type}:
\begin{equation}
R_\Lambda^\Phi=Z_\Lambda^\Phi \Lambda^2 ,
\label{mass-reg}
\end{equation}
where $Z_\Lambda^\Phi$ denotes the \textbf{wave function normalization}  constant for the field $\Phi$. 
We discuss later (in the end of section IV) whether the result is independent of the choice of the infrared regulator or not.

For the infrared regulator of the mass type, thus the flow equation reads
\begin{align}
% \nonumber\\&
\partial_{t} \Gamma_{\Lambda}  
%\nonumber\\
 =&  \frac{NgH}{(4\pi)^{\frac{D}{2}}}  \Big\{  ( W_{\Lambda}^{\prime} )^{1- \frac{D}{2}}  ( 2 - \eta_{\Lambda} ) Z_{\Lambda} \Lambda^{2}  
\nonumber\\&  
\times  \int_{0}^{\infty} d \tau \tau^{1 - \frac{D}{2}} e^{- \tau Z_{\Lambda} \Lambda^{2}}   
\nonumber\\&  \times
\frac{ (D-2)e^{-\tau W_{\Lambda}^{\prime}gH}+e^{-3\tau W_{\Lambda}^{\prime}gH}+e^{\tau W_{\Lambda}^{\prime}gH}}{1-e^{-2\tau W_{\Lambda}^{\prime}gH}}   
\nonumber\\&
-   ( W_{\Lambda}^{\prime} )^{1- \frac{D}{2} }  ( 2 - \eta_{\Lambda} ) Z_{\Lambda}  \Lambda^{2}  
\nonumber\\& \times
\int_{0}^{\infty} d \tau \tau^{1 - \frac{D}{2}}  e^{- \tau Z_{\Lambda} \Lambda^{2}}  \frac{  e^{-\tau W_{\Lambda}^{\prime} gH} }{1-e^{-2\tau W_{\Lambda}^{\prime} gH}} 
\nonumber\\&
+   \alpha_\Lambda^{ \frac{D}{2}-1}  ( 2 - \eta_{\Lambda} ) Z_{\Lambda}  \Lambda^{2}  
\nonumber\\& \times
\int_{0}^{\infty} d \tau \tau^{1 - \frac{D}{2}}  e^{- \tau Z_{\Lambda} \Lambda^{2}}  \frac{  e^{-\tau \alpha_\Lambda^{-1} gH} }{1-e^{-2\tau \alpha_\Lambda^{-1} gH}} 
\nonumber\\&
 - (\tilde{Z}_{\Lambda})^{1-\frac{D}{2}} (2 - \tilde{\eta}_\Lambda) \tilde{Z}_{\Lambda} \Lambda^{2} 
\nonumber\\& \times
\int_{0}^{\infty} d \tau \tau^{1 - \frac{D}{2}}  e^{- \tau \tilde{Z}_{\Lambda} \Lambda^{2}}  \frac{2e^{-\tau \tilde{Z}_{\Lambda} gH}}{1-e^{-2\tau \tilde{Z}_{\Lambda} gH}}  \Big\} ,
\label{flow-eq1}
\end{align}
where we have introduced  the \textbf{anomalous dimensions}:
\begin{align}
\eta_\Lambda &:=-\partial_t \ln Z_\Lambda =-Z_\Lambda^{-1} \partial_t Z_\Lambda , 
\nonumber\\  
\Tilde{\eta}_\Lambda &:=-\partial_t \ln \Tilde{Z}_\Lambda =-\Tilde{Z}_\Lambda^{-1} \partial_t \Tilde{Z}_\Lambda.
\end{align}

We find that the integral with respect to $\tau$ on the right-hand side of the flow equation  is divergent at $D=4$ in the $\tau=0$ region, which is an \textbf{ultraviolet divergence}. 
This divergence is independent of the \textbf{infrared  divergence} coming from $\tau=\infty$ region due to the factor $e^{\tau W_{\Lambda}^{\prime}gH}$ for which the Nielsen-Olesen instability is responsible.
This ultraviolet divergence is due to the fact that  the momentum-independent infrared cutoff function of the mass type  does not suppress the high-momentum modes.
This aspect is a short-coming of the mass-type infrared regulator. Other choices of the infrared regulator are discussed in the end of this section.

Therefore, we  first remove the ultraviolet divergence at $\tau=0$.
This is done in the standard way  by adopting the minimal subtraction, i.e., $\overline{MS}$ scheme. 
See Appendix B for the details. 
Thus we arrive at the flow equation without the ultraviolet divergence:
\begin{align}
\partial_{t} \Gamma_{\Lambda}   
=&  \frac{N}{2} \frac{2gH }{(4\pi)^{2 }} 
  \left(  -\ln \frac{2gH}{4 \pi \mu^2 } - \gamma \right) 
\Biggr\{ ( W_{\Lambda}^{\prime} )^{-1}  ( 2 - \eta_{\Lambda} ) Z_{\Lambda} \Lambda^{2}  
\nonumber\\&  
\times  %\nonumber\\ & \times
    \Bigg[   \zeta \left( 0, \frac{1}{2} +\frac{Z_{\Lambda} \Lambda^{2}}{2 W_{\Lambda}^{\prime}gH}  \right) 
    +\zeta \left( 0, \frac{3}{2} +\frac{Z_{\Lambda} \Lambda^{2}}{2 W_{\Lambda}^{\prime}gH}  \right) 
\nonumber\\&      + \zeta \left( 0, -\frac{1}{2} +\frac{Z_{\Lambda} \Lambda^{2}}{2 W_{\Lambda}^{\prime}gH}  \right)   \Bigg] 
%+ \epsilon \zeta^{(1,0)}\left( 0, \frac{1}{2} +\frac{Z_{\Lambda} \Lambda^{2}}{2 W_{\Lambda}^{\prime}gH}  \right)  
\nonumber\\ &
   + \alpha_\Lambda   ( 2 - \eta_{\Lambda} ) Z_{\Lambda}  \Lambda^{2}     \zeta \left( 0, \frac{1}{2} +\frac{Z_{\Lambda} \Lambda^{2}}{2 \alpha_\Lambda^{-1} gH}  \right)
%+ \epsilon \zeta^{(1,0)}\left( 0, \frac{3}{2} +\frac{Z_{\Lambda} \Lambda^{2}}{2 W_{\Lambda}^{\prime}gH}  \right)  
\nonumber\\& 
 - 2 (2 - \tilde{\eta}_\Lambda) \Lambda^{2}  \zeta \left( 0, \frac{1}{2} +\frac{ \Lambda^{2}}{2 gH}  \right)   \Biggr\}
\nonumber\\ & 
 +  \frac{N}{2} \frac{2gH }{(4\pi)^{2 }} 
 \Biggr\{  (W_{\Lambda}^{\prime} )^{-1}  ( 2 - \eta_{\Lambda} ) Z_{\Lambda} \Lambda^{2}  
\nonumber\\&  
 %\nonumber\\ & \times
    \times  \Bigg[   \zeta^{(1,0)} \left( 0, \frac{1}{2} +\frac{Z_{\Lambda} \Lambda^{2}}{2 W_{\Lambda}^{\prime}gH}  \right) 
\nonumber\\&     \quad\quad
 +\zeta^{(1,0)} \left( 0, \frac{3}{2} +\frac{Z_{\Lambda} \Lambda^{2}}{2 W_{\Lambda}^{\prime}gH}  \right)      
%+ \epsilon \zeta^{(1,0)}\left( 0, -\frac{1}{2} +\frac{Z_{\Lambda} \Lambda^{2}}{2 W_{\Lambda}^{\prime}gH}  \right)  
\nonumber\\ &
 \quad\quad  + \zeta^{(1,0)} \left( 0, -\frac{1}{2} +\frac{Z_{\Lambda} \Lambda^{2}}{2 W_{\Lambda}^{\prime}gH}  \right)   
\nonumber\\&  \quad\quad
-2 \zeta \left( 0, \frac{1}{2} +\frac{Z_{\Lambda} \Lambda^{2}}{2 W_{\Lambda}^{\prime}gH}  \right)
 \Bigg]   
\nonumber\\&
 \quad\quad +   \alpha_\Lambda   ( 2 - \eta_{\Lambda} ) Z_{\Lambda}  \Lambda^{2}    \zeta^{(1,0)}\left( 0, \frac{1}{2} +\frac{Z_{\Lambda} \Lambda^{2}}{2 \alpha_\Lambda^{-1} gH}  \right)  
\nonumber\\&
 \quad   - 2   (2 - \tilde{\eta}_\Lambda)   \Lambda^{2}   \zeta^{(1,0)}\left( 0, \frac{1}{2} +\frac{ \Lambda^{2}}{2 gH}  \right)     \Biggr\} ,
\label{flow-eq5}
\end{align}
where  $\zeta (z, \lambda)$ is the  \textbf{generalized Riemann $\zeta$-function} or the \textbf{Hurwitz $\zeta$-function} defined by (\ref{zeta-def}) and its integral representation (\ref{integral-rep-zeta}), and its derivatives $\zeta^{(m,n)}(z, \lambda)$ are defined by (\ref{zeta-def-derivative}).

%%%%%%%%%%%%%%%%%%%%%%%%%%%%%%%%%%%%%%%%%%%%%%%%%

\section{Absence of the instability}

%%%%%%%%%%%%%%%%%%%%%%%%%%%%%%%%%%%%%%%%%%%%%%%%%

For large $\Lambda$, we can take the approximation:
\begin{align}
\text{i}) & \   W_\Lambda (\Theta)  = \Theta \Rightarrow W^\prime_\Lambda (\Theta) \equiv 1 %W''_\Lambda (\theta) \equiv 0, \dots  
 \nonumber\\  & \Longleftrightarrow Z_\Lambda \equiv 1 \Rightarrow \eta_\Lambda \equiv 0 , 
 \nonumber\\
 \text{ii)} & \  \Tilde{Z}_\Lambda \equiv 1 \Rightarrow  \Tilde{\eta}_\Lambda \equiv 0 ,
 \nonumber\\
 \text{iii)} & \  \alpha_\Lambda \equiv\alpha_{\Lambda_{\rm UV}} = \text{const.} \ge 0 .
\end{align}
This is a good approximation for $\Gamma_\Lambda$ at sufficiently large $\Lambda$. 
If we choose $\alpha_\Lambda^{-1} \equiv W^\prime_\Lambda $, the second and  third terms cancel on the right-hand side of the flow equation (\ref{flow-eq1}), which corresponds to the Feynman gauge. 
If we choose $\alpha_\Lambda \equiv 0$, the  third term on the right-hand side of the flow equation (\ref{flow-eq1}) vanishes, which corresponds to the Landau gauge. 

Then we obtain an approximate flow equation for   large $\Lambda$:
\begin{align}
\partial_{t} \Gamma_{\Lambda}   
=&  \frac{N}{2} \frac{2gH }{(4\pi)^{2 }} 
  \left( - \ln \frac{2gH}{4 \pi \mu^2 } - \gamma \right)  2   \Lambda^{2}  
\nonumber\\&  
\times  %\nonumber\\ & \times
   \Biggr\{   
 \zeta \left( 0, \frac{3}{2} +\frac{  \Lambda^{2}}{2  gH}  \right) 
    + \zeta \left( 0, -\frac{1}{2} +\frac{ \Lambda^{2}}{2  gH}  \right)   
%+ \epsilon \zeta^{(1,0)}\left( 0, \frac{1}{2} +\frac{Z_{\Lambda} \Lambda^{2}}{2 W_{\Lambda}^{\prime}gH}  \right)  
\nonumber\\ &
 -    \zeta \left( 0, \frac{1}{2} +\frac{ \Lambda^{2}}{2 gH}  \right)   
+ \alpha_\Lambda       \zeta \left( 0, \frac{1}{2} +\frac{  \Lambda^{2}}{2 \alpha_\Lambda^{-1} gH}  \right)
%+ \epsilon \zeta^{(1,0)}\left( 0, \frac{3}{2} +\frac{Z_{\Lambda} \Lambda^{2}}{2 W_{\Lambda}^{\prime}gH}  \right)  
   \Biggr\}
\nonumber\\ &
 +  \frac{N}{2} \frac{2gH }{(4\pi)^{2 }}   2   \Lambda^{2}  
 \Biggr\{  
         \zeta^{(1,0)} \left( 0, \frac{3}{2} +\frac{ \Lambda^{2}}{2  gH}  \right)      
\nonumber\\&   \quad\quad\quad\quad\quad\quad\quad
+ \zeta^{(1,0)} \left( 0, -\frac{1}{2} +\frac{  \Lambda^{2}}{2  gH}  \right)   %+ \epsilon \zeta^{(1,0)}\left( 0, -\frac{1}{2} +\frac{Z_{\Lambda} \Lambda^{2}}{2 W_{\Lambda}^{\prime}gH}  \right)  
\nonumber\\ & \quad\quad\quad\quad\quad\quad\quad 
-2 \zeta \left( 0, \frac{1}{2} +\frac{ \Lambda^{2}}{2 gH}  \right) 
\nonumber\\&  \quad\quad\quad\quad\quad\quad\quad 
   -       \zeta^{(1,0)}\left( 0, \frac{1}{2} +\frac{ \Lambda^{2}}{2 gH}  \right) 
\nonumber\\&   \quad\quad\quad\quad\quad\quad\quad 
+   \alpha_\Lambda       \zeta^{(1,0)}\left( 0, \frac{1}{2} +\frac{ \Lambda^{2}}{2 \alpha_\Lambda^{-1} gH}  \right)  
   \Biggr\} .
\label{flow-eq6}
\end{align}
Then the flow equation can be cast into the total derivative form:
\begin{align}
\partial_{t} \Gamma_{\Lambda}   
=&  \partial_{t}   \Biggr\{  \frac{N}{2} \frac{(2gH)^2 }{(4\pi)^{2 }} 
  \left(  \ln \frac{2gH}{4 \pi \mu^2 } + \gamma \right)   
\nonumber\\&  
\times  %\nonumber\\ & \times
   \Biggr[   
 \zeta \left( -1, \frac{3}{2} +\frac{  \Lambda^{2}}{2  gH}  \right) 
    + \zeta \left( -1, -\frac{1}{2} +\frac{ \Lambda^{2}}{2  gH}  \right)   
%+ \epsilon \zeta^{(1,0)}\left( 0, \frac{1}{2} +\frac{Z_{\Lambda} \Lambda^{2}}{2 W_{\Lambda}^{\prime}gH}  \right)  
\nonumber\\ &
 -    \zeta \left( -1, \frac{1}{2} +\frac{ \Lambda^{2}}{2 gH}  \right)   + \alpha_\Lambda       \zeta \left( -1, \frac{1}{2} +\frac{  \Lambda^{2}}{2 \alpha_\Lambda^{-1} gH}  \right)
%+ \epsilon \zeta^{(1,0)}\left( 0, \frac{3}{2} +\frac{Z_{\Lambda} \Lambda^{2}}{2 W_{\Lambda}^{\prime}gH}  \right)  
   \Biggr]
\nonumber\\ &
 -  \frac{N}{2} \frac{(2gH)^2 }{(4\pi)^{2 }}   
 \Biggr[  
         \zeta^{(1,0)} \left( -1, \frac{3}{2} +\frac{ \Lambda^{2}}{2  gH}  \right)      
\nonumber\\&    \quad\quad\quad\quad\quad\quad
+ \zeta^{(1,0)} \left( -1, -\frac{1}{2} +\frac{  \Lambda^{2}}{2  gH}  \right)   %+ \epsilon \zeta^{(1,0)}\left( 0, -\frac{1}{2} +\frac{Z_{\Lambda} \Lambda^{2}}{2 W_{\Lambda}^{\prime}gH}  \right)  
\nonumber\\ &  \quad\quad\quad\quad\quad\quad
-2 \zeta \left( -1, \frac{1}{2} +\frac{ \Lambda^{2}}{2 gH}  \right) 
\nonumber\\&  \quad\quad\quad\quad\quad\quad 
   -       \zeta^{(1,0)}\left( -1, \frac{1}{2} +\frac{ \Lambda^{2}}{2 gH}  \right) 
\nonumber\\&  \quad\quad\quad\quad\quad\quad
+   \alpha_\Lambda       \zeta^{(1,0)}\left( -1, \frac{1}{2} +\frac{ \Lambda^{2}}{2 \alpha_\Lambda^{-1} gH}  \right)  \Biggr]
   \Biggr\}  ,
\end{align}
where 
we have used the relation following from the definition: 
\begin{equation}
  \zeta^{(m,1)} (z-1, \lambda) = \frac{\partial}{\partial \lambda} \zeta^{(m,0)} (z-1, \lambda) = - \zeta^{(m,0)} (z, \lambda) ,
\end{equation}
which yields 
\begin{align}
 & \partial_t \zeta^{(m,0)}\left( -1, a +\frac{ \Lambda^{2}}{2 gH}  \right)  
\nonumber\\    
=&  \zeta^{(m,1)}\left( -1 , a +\frac{ \Lambda^{2}}{2 gH}  \right) \frac{ \Lambda^{2}}{gH} 
\nonumber\\    
=&  - \zeta^{(m,0)}\left( 0 , a +\frac{ \Lambda^{2}}{2 gH}  \right) \frac{ \Lambda^{2}}{gH}  .
\end{align}

We take into account the fact that the effective average action  $\Gamma_{\Lambda}$ at $\Lambda=\Lambda_{\rm UV}=\infty$ is given by the bare action for the classical chromomagnetic  field   background:
\begin{equation}
 \Gamma_{\Lambda=\infty} =\frac{1}{4} \left( \mathscr{F}_{\mu \nu}^A [\mathscr{V}] \right)^2 =\frac{1}{2} H^2 .
\end{equation}
Then an approximate solution is obtained by integrating the flow equation from $\Lambda=\Lambda_{\rm UV}=\infty$ to $\Lambda$:
\begin{align}
 \Gamma_\Lambda  =& \frac{1}{2}H^2+\Tilde{V}_\Lambda (H) ,
\end{align}
with 
\begin{align}
\Tilde{V}_\Lambda (H)   =&  - \frac{N}{2} \frac{(2gH)^2 }{(4\pi)^{2 }} 
  \left(  \ln \frac{2gH}{4 \pi \mu^2 } + \gamma \right)   
\nonumber\\&  
\times  %\nonumber\\ & \times
   \Biggr[   
 \zeta \left( -1, \frac{3}{2} +\frac{  \Lambda^{2}}{2  gH}  \right) 
    + \zeta \left( -1, -\frac{1}{2} +\frac{ \Lambda^{2}}{2  gH}  \right)   
%+ \epsilon \zeta^{(1,0)}\left( 0, \frac{1}{2} +\frac{Z_{\Lambda} \Lambda^{2}}{2 W_{\Lambda}^{\prime}gH}  \right)  
\nonumber\\ &
 -    \zeta \left( -1, \frac{1}{2} +\frac{ \Lambda^{2}}{2 gH}  \right)   + \alpha_\Lambda       \zeta \left( -1, \frac{1}{2} +\frac{  \Lambda^{2}}{2 \alpha_\Lambda^{-1} gH}  \right)
%+ \epsilon \zeta^{(1,0)}\left( 0, \frac{3}{2} +\frac{Z_{\Lambda} \Lambda^{2}}{2 W_{\Lambda}^{\prime}gH}  \right)  
   \Biggr]
\nonumber\\ &
 +  \frac{N}{2} \frac{(2gH)^2 }{(4\pi)^{2 }}   
 \Biggr[  
         \zeta^{(1,0)} \left( -1, \frac{3}{2} +\frac{ \Lambda^{2}}{2  gH}  \right)      
\nonumber\\&  \quad\quad\quad\quad 
+ \zeta^{(1,0)} \left( -1, -\frac{1}{2} +\frac{  \Lambda^{2}}{2  gH}  \right)   %+ \epsilon \zeta^{(1,0)}\left( 0, -\frac{1}{2} +\frac{Z_{\Lambda} \Lambda^{2}}{2 W_{\Lambda}^{\prime}gH}  \right)  
\nonumber\\&  \quad\quad\quad\quad 
   -       \zeta^{(1,0)}\left( -1, \frac{1}{2} +\frac{ \Lambda^{2}}{2 gH}  \right) 
\nonumber\\ &  \quad\quad\quad\quad 
-2 \zeta \left( -1, \frac{1}{2} +\frac{ \Lambda^{2}}{2 gH}  \right) 
\nonumber\\&  \quad\quad\quad\quad 
+   \alpha_\Lambda       \zeta^{(1,0)}\left( -1, \frac{1}{2} +\frac{ \Lambda^{2}}{2 \alpha_\Lambda^{-1} gH}  \right)  \Biggr] ,
\end{align}
where $\Tilde{V}_\Lambda (H)=0$ at $\Lambda=\infty$.

Using the formula \cite{EORBZ94,Erdelyi81,WW27,AS72,GR72}:
\begin{equation}
\zeta (-1, \lambda) =-\frac{1}{2} \lambda^2 +\frac{1}{2} \lambda -\frac{1}{12}
%= -\frac{1}{2} \left(\lambda-\frac{1}{2} \right)^2 + \frac{1}{24} 
\ , \  (\lambda \in \mathbb{R}) ,
\end{equation}
we find that the first term proportional to $\ln \frac{gH}{\mu^2}$ is real valued for $\frac{gH}{\mu^2}>0$, since
\begin{equation}
\zeta \left(-1, \frac{3}{2}+\frac{r}{2} \right) + \zeta \left(-1, -\frac{1}{2}+\frac{r}{2} \right) = -\frac{11}{12}-\frac{1}{4} r^2  .
% \ r :=  \frac{\Lambda^2}{gH} .
\end{equation}
On the other hand, the recursion relation \cite{EORBZ94}:
%\begin{equation}
%\zeta^{(1,0)} (-1, a+n)=\zeta^{(1,0)} (-1, a)+ \sum \limits_{n=0}^{n-1} (k+a) \ln (k+a)  ,
%\end{equation}
%in particular, 
\begin{equation}
\zeta^{(1,0)} (-1, a+1) = \zeta^{(1,0)} (-1, a) + a \ln a ,
\end{equation}
leads to
\begin{align}
&\zeta^{(1,0)} \left(-1, \frac{3}{2}+\frac{r}{2} \right) +\zeta^{(1,0)} \left(-1, -\frac{1}{2}+\frac{r}{2}  \right) \notag \\
&=2\zeta^{(1,0)} \left(-1, \frac{1}{2}+\frac{r}{2} \right)
\nonumber\\&
+\frac{1+r}{2} \ln \frac{1+r}{2}-\frac{-1+r}{2} \ln \frac{-1+r}{2}.
\end{align}
Note that $\zeta^{(1,0)}(-1,\lambda)$ is real valued for $\lambda>0$. 
See Appendix B. 
Thus, we arrive at the effective potential for large $\Lambda$, e.g., in the case of $\alpha_\Lambda \equiv 1$:
\begin{align}
V_\Lambda (H)  &=  \frac{1}{2} H^2 +\frac{1}{16\pi^2} \Lambda^2 \left[ \ln \frac{gH}{\mu^2} +\frac{1}{4}-C \right] 
\notag \\ &
- \frac{2}{16\pi^2} gH \Lambda^2 \ln \frac{\Lambda^2-gH}{\Lambda^2+gH} \notag \\
&+ \frac{1}{16\pi^2} g^2 H^2 \Bigg[ \frac{11}{3} \ln \frac{gH}{\mu^2} 
\nonumber\\ & 
+ 2 \ln  \frac{\Lambda^2 +gH}{gH}  +  2 \ln \frac{\Lambda^2 -gH}{gH}   
 \notag \\ &
  -\frac{11}{3}C -4 \ln 2 -\frac{1}{3} +8 \zeta^{(1,0)} \left( -1, \frac{1}{2}+ \frac{\Lambda^2}{2gH} \right) \Bigg ] ,
\label{solution}
\end{align}
while 
in the case of $\alpha_\Lambda \equiv 0$:
\begin{align}
V_\Lambda (H)  &=  \frac{1}{2} H^2 +\frac{1}{16\pi^2} \Lambda^2 \left[ \frac{1}{2} \ln \frac{gH}{\mu^2} +\frac{1}{4}- \frac{1}{2}C \right] \notag \\
&- \frac{2}{16\pi^2} gH \Lambda^2 \ln \frac{\Lambda^2-gH}{\Lambda^2+gH} \notag \\
&+ \frac{1}{16\pi^2} g^2 H^2 \Bigg[ -\frac{1}{6} \ln \frac{gH}{\mu^2} 
\nonumber\\ & + 2 \ln  \frac{\Lambda^2 +gH}{gH}  +  2 \ln \frac{\Lambda^2 -gH}{gH}   
 \notag \\
&  -\frac{23}{6}C -4 \ln 2 -\frac{1}{3} +4 \zeta^{(1,0)} \left( -1, \frac{1}{2}+ \frac{\Lambda^2}{2gH} \right) \Bigg ] .
\label{solution-Landau}
\end{align}
The same effective potential is obtained  by solving the flow equation (\ref{flow-eq1}) to obtain the effective potential with the ultraviolet divergence  and then removing the ultraviolet divergence by the same method as that above. 
See Appendix B for more details for the Hurwitz $\zeta$-function. 

It is instructive to give a comment on the gauge parameter. 
In the Lorenz gauge there is a privileged choice: $\alpha=0$ is a fixed point. 
Whereas there is no special choice for the gauge parameter in the MA gauge: there is no fixed point for $\alpha$ at least in the one loop level. 
See Appendix C.

%-------- Figure ----------------------------
\begin{figure}[t]
\begin{center}
\includegraphics[scale=0.35]{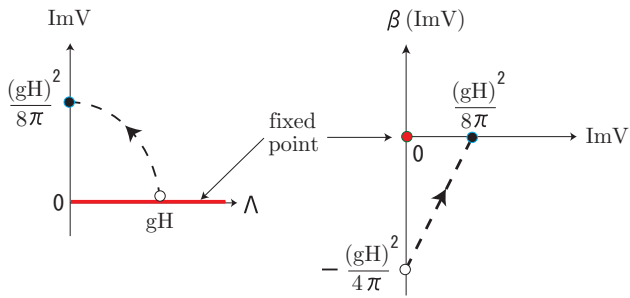}
\quad\quad
\end{center}
\vskip -0.5cm
\caption{ 
(Left panel) Imaginary part $\text{Im} V_\Lambda$ of an approximate average effective potential as a function of $\Lambda$,
(Right panel)    $\beta$ function as a function of $\text{Im} V_\Lambda$.
Here $\text{Im} V_\Lambda \equiv 0$ corresponds to a fixed point of the functional renormalization group. 
}
\label{fig:RG-Savvidy} 
\end{figure}
%---------------------------------------------

For the large $\Lambda$ satisfying  $\Lambda^2 \ge gH$, $\Tilde{V}_\Lambda (H)$ is real valued and $V_\Lambda (H)$ has no imaginary part:
\begin{equation}
\text{Im} V_\Lambda (H) =0  \ \text{for} \ \Lambda^2 \ge gH ,
\end{equation}
and
\begin{equation}
  \partial_t \text{Im} V_\Lambda (H) = 0 \ \text{for} \ \Lambda^2 \ge gH  .
\end{equation}
Therefore, the Nielsen-Olesen instability disappears for any value of $\Lambda$, in particular even at $\Lambda=0$ according to the above argument of the fixed point for the pure imaginary part of the flow equation. 
See Fig.~\ref{fig:RG-Savvidy}.

For the small $\Lambda$ satisfying $\Lambda^2 <gH$, however, the effective average potential $V_\Lambda (H)$ obtained above has the non-vanishing imaginary part:
\begin{align}
  \text{Im} V_\Lambda (H) &= \frac{4}{16\pi^2} g^2 H^2 \frac{\frac{\Lambda^2}{gH}-1}{2} \ln (-1)/i 
\nonumber\\ & =\frac{1}{8\pi} gH (gH-\Lambda^2)    \ \ \text{for} \ \Lambda^2 <gH ,
\end{align}
which yields the nontrivial flow of the imaginary part:
\begin{equation}
\partial_t \text{Im} V_\Lambda (H) = - \frac{1}{4\pi} gH \Lambda^2 < 0 \ \ \text{for} \ \Lambda^2<gH .
\end{equation}
The $\beta$ function for $\text{Im} V_\Lambda$ is obtained as
\begin{equation}
 \beta(   \text{Im} V_\Lambda ) = 2 \text{Im} V_\Lambda  -  \frac{1}{4\pi} (gH)^2  < 0 \ \ \text{for} \ \Lambda^2<gH.
\end{equation}
This is not a contradiction, since the approximate solution of $V_\Lambda (H)$ obtained above is not considered to be valid in the small $\Lambda$ region; $\Lambda^2 <gH$. 
In fact, the derivative $\partial_t \text{Im} V_\Lambda (H)$ has the discontinuity at $\Lambda^2=gH$.
The effective potentials obtained above  reproduce the Nielsen-Olesen result by putting $\Lambda = 0$:
\begin{align}
%\hspace*{-100mm}
V_{\rm NO}(H)
 =\frac12 H^2
 %\left[1 
 +\frac{\frac{22}{3}}{16\pi^2}g^2 \frac12 H^2
   \left(\ln\frac{gH}{\mu^2} + c  \right)
% \right]
+  i \ \frac{g^2 H^2}{8\pi}  .
\end{align}

The above argument for the \textbf{absence of the instability} or the \textbf{vanishing of the imaginary part in the average effective potential} $V_\Lambda (H)$ was done for a specific choice of the infrared regulator. However, the result will be true for any other choice of the infrared regulator, since the infrared regulator $R_\Lambda (p^2)$ is constructed in such a way that any infrared cutoff function approaches the   asymptotic form of the same form as the mass-type one in the large $\Lambda$:
\begin{equation}
R_\Lambda^\Phi (p^2) \to Z_\Lambda^\Phi \Lambda^2 \ \ \text{for} \ \Lambda^2 \to \infty.
\end{equation}
Indeed, this condition must be imposed to reproduce the ``one-loop result'' in the large $\Lambda$, which is indeed one of the properties required to hold for the infrared regulator \cite{LP99}.

The claim can be explicitly checked 
for the infrared regulators, e.g.,  the optimal type \cite{Litim00},
\begin{equation}
R_\Lambda (p^2) = Z_\Lambda (\Lambda^2-p^2) \theta (\Lambda^2-p^2) ,
\end{equation}
and   the step function,
\begin{equation}
R_\Lambda (p^2) =Z_\Lambda \Lambda^2 \theta (\Lambda^2-p^2) .
\end{equation}
This is nontrivial  for the exponential-type,
\begin{equation}
R_\Lambda (p^2) =   \frac{ p^2 }{e^{\frac{p^2}{Z_{\Lambda} \Lambda^{2}}} - 1}
= \frac{ p^2 e^{-\frac{p^2}{Z_{\Lambda} \Lambda^{2} }}}{1-e^{-\frac{p^2}{Z_{\Lambda} \Lambda^{2}}}}
 ,
%= Z_{\Lambda} \Lambda^{2} \frac{y}{e^{y} - 1}  , 
%\ y := \frac{u}{Z_{\Lambda} \Lambda^{2}} .
\end{equation}
since the momentum integration is difficult to be performed explicitly for this choice. 

Moreover, it is important to confirm the statement explicitly for the choice of the   infrared regulator $R_\Lambda(\Gamma^{(2)}(p^2))$ with the nontrivial argument $\Gamma^{(2)}(p^2)$ proposed in \cite{RW94}, since it is demonstrated in \cite{Gies02} that such a  choice of the argument for the infrared regulator  is actually essential to control the physical limit $\Lambda \to 0$.

Here is the good place to review the preceding works on which this work is based. 
In the paper by Reuter and Wetterich \cite{RW94},  a new nonperturbative flow equation for the average effective action was proposed for Yang-Mills theories. 
The subsequent works \cite{RW97,Gies02,EGP11} are more or less based on this framework. 
In  a subsequent work by them \cite{RW97}, it was applied to the calculation of the gluon condensation and the computation of the effective action for a uniform chromomagnetic field to examine the instability of the Savvidy vacuum. 
This work improved the earlier error in the evaluation of the flow equation in preceding work, but it didn't find a desired gluon condensation in a simple way, since the strong infrared effects were cut off in  an \textit{ad hoc} way  by introducing  effectively an infrared fixed point by hand.  Therefore, the resulting flow equation taken at face value shows a Landau-pole-type singularity.  

The work by Gies \cite{Gies02} is an improvement of the earlier works by Reuter and Wetterich, which was called ``spectrally adjusted'' RG flow or spectral adjustment of the RG procedure. 
Gies has succeeded to estimate the effect of the  $\partial_\Lambda \Gamma_\Lambda^{(2)}$ terms coming from the argument of the infrared regulator function $R_\Lambda(\Gamma_\Lambda^{(2)})$ that had been dropped in the preceding work. 
These terms  become essential when the RG flow rapidly changes in the strong coupling domain.
In fact, this improvement is necessary to derive the infrared fixed point, namely, the running coupling constant reaching  a finite and nonzero value  in the limit  $\Lambda \to 0$ without encountering divergence. 

In \cite{RW97} and \cite{Gies02}, an ansatz  of the power series $W_\Lambda(\Theta)=\sum_{n=0}^{\infty} \frac{1}{n!} w_n(\Lambda) \Theta^n$ is adopted   for $W_\Lambda(\Theta)$,  not to solve the flow, but to define the running coupling constant from the coefficient $w_1(\Lambda)$ in front of the term  $\Theta :=\frac14 F_{\mu\nu}^2$.  It should be remarked that different choices for the definition of the running coupling can lead to different results, since the running coupling itself is not meaningful quantity intrinsically in the sense that  it depends on the scheme and the definition. 

In the works \cite{RW97,Gies02}, the magnetic field is only used as a technical tool to determine the flow equation. One need not assume that there is a physical magnetic background field. The same results for the running coupling would be obtained with, e.g., a heat-kernel expansion of the traces that is blind to the instability. Therefore, the NO instability is not an issue at all in these works.
Still, calculating the flow using the magnetic field as a tool, of course, contains contributions from the Nielsen-Olesen mode, as it is also true for the one-loop calculation.

In the work by Eichhorn, Gies and Pawlowski \cite{EGP11}, on the other hand, the full propagators were used to compute the gluon condensate.
The negative eigenvalues of the spin-1  Laplacian can potentially botch the computations. Therefore, they have used the self-dual background in order to avoid these complications from the beginning.

%(Also at finite T with Jens Braun, the instability can be more dangerous and we needed to control  it).

%%%%%%%%%%%%%%%%%%%%%%%%%%%%%%%%%%%%%%%%%%%%%%%%%
 
\section{gluon mass generation and vacuum condensations}

%%%%%%%%%%%%%%%%%%%%%%%%%%%%%%%%%%%%%%%%%%%%%%%%%

 The above approximate solution (\ref{solution}) eventually has the imaginary part and hence cannot be used in the limit $\Lambda \to 0$. 
As will be shown in this section, however, the approximate solution obtained in the same type of approximations has the limit $\Lambda \to 0$ without developing the imaginary part, if the effects of mass generation are incorporated into the analysis. 
Such mass generation is expected to occur, as established in the numerical simulations on the lattice \cite{AS99,BCGMP03,CDG-numerical}. 
\footnote{
This means the mass generation for the off-diagonal gluons in the MA gauge. 
For the diagonal gluon, this is not yet confirmed even for the MA gauge. 
}

We introduce  the mixed composite operators of gluons and ghosts: 
For $SU(2)$,  
\begin{equation}
\mathcal{O}=\frac{1}{2} A_\mu^a A^{\mu a} +\alpha i \Bar{C}^a C^a \ \ (a=1, 2).
\end{equation}
We then study the mass generation for the off-diagonal gluons (and ghosts), originating from the dimension-two condensation $\langle \mathcal{O} \rangle$.
It is shown \cite{Kondo01} that the dimension-two condensation $\langle \mathcal{O} \rangle$ is BRST invariant % 
\footnote{
We can construct a gauge-invariant version of the composite operator of mass dimension-two, see \cite{KMS05,KMS06,Kondo06}.
} 
 in the modified MA gauge  \cite{Kondo98} defined by  the GF+FP term:%
\footnote{
In the MA gauge, the four-point interaction $AA\bar C C$ appears irrespective of the gauge-fixing parameter $\alpha$ and it generates the four-point ghost self-interaction $\bar C C\bar C C$ by quantum corrections. 
Therefore, such a four-point ghost self-interaction  is indispensable to maintain the renormalizability.   
The naive MA gauge is nonrenormalizable, since it does not include the four-point ghost self-interactions. 
In the modified MA gauge, the strength of the four-point ghost self-interactions is proportional to the gauge-fixing parameter $\alpha$.
Such a four-point ghost self-interaction  follows from   the $OSp(D,2)$ invariance. 
See \cite{Ferrari13} for the meaning of the four ghost interactions in the MA gauge. 
}
\begin{align}
\mathscr{L}_{\rm GF+FP}^{\rm MA}  
=& N^{a} F^{a} + \frac{\alpha}{2} N^{a} N^{a}  
\nonumber\\&
+ i \bar{C}^{a} \mathscr{D}_{\mu}^{ab} [V] \mathscr{D}_{\mu}^{ bc} [V] C^{c} 
\nonumber\\ &
- g^{2} \epsilon^{ab} \epsilon^{cd} i \bar{C}^{a} C^{d} A_{\mu}^{c} A_{\mu}^{ b} 
\nonumber\\
&+ g  i \bar{C}^{a} \epsilon^{ab} ( \mathscr{D}_{\mu}^{bc} [V] A_{\mu}^{ c} ) C^{3} 
\nonumber\\&
+ \alpha g \epsilon^{ab} i \bar{C}^{a} N^{b} C^{3} + \frac{\alpha}{4} g^{2} \epsilon^{ab} \epsilon^{cd} \bar{C}^{a} \bar{C}^{b} C^{c} C^{d} , 
\end{align}
which is deduced from the $OSp(D,2)$-invariant form:
\begin{align}
\mathscr{L}_{\rm GF+FP}^{\rm MA} 
=&  i \bm{\delta} \bar{\bm{\delta}} \left( \frac{1}{2}  A_{\mu}^{a} A_{\mu}^{ a} + \frac{\alpha}{2} i \bar{C}^{a} C^{a} \right) 
\nonumber\\ 
=&  - i \bm{\delta} \biggl[ \bar{C}^{a} \left( F^{a} + \frac{\alpha}{2} N^{a} \right) - \frac{\alpha}{2} g i \bar{C}^{a} \epsilon^{ab3} \bar{C}^{b} C^{3} \biggr] , 
\nonumber\\ &
 F^{a} := \mathscr{D}_\mu^{ab} [V] A_\mu^b ,
\end{align}
where $\bm{\delta}$ and $\bar{\bm{\delta}}$ are, respectively, the BRST and anti-BRST transformations.

%-------- Figure ----------------------------
\begin{figure}[t]
\begin{center}
\includegraphics[scale=0.25]{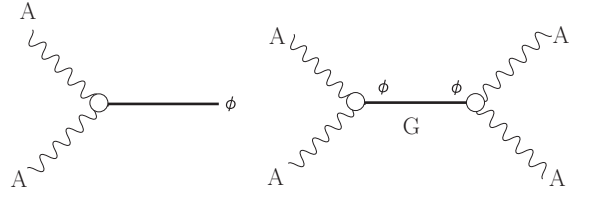}
\quad\quad
\end{center}
\vskip -0.5cm
\caption{ 
(Left panel) Vertex joining the collective field $\phi$ to two off-diagonal gluon fields $A$,
(Right panel) an exchange of the collective field $\phi$.
}
\label{fig:composite} 
\end{figure}
%---------------------------------------------

According to \cite{EW94}, we introduce a new field $\phi$ which is an auxiliary field  with no kinetic term represented by the Lagrangian density,
\begin{align}
\mathscr{L}_\phi
&=\frac{1}{2} \left( \phi+G\mathcal{O} \right)^\dagger G^{-1} \left( \phi +G\mathcal{O} \right) \nonumber\\
&=\frac{1}{2} \phi^\dagger G^{-1} \phi +\phi^\dagger \mathcal{O} +\frac{1}{2} \mathcal{O}^\dagger G \mathcal{O} ,
\end{align}
by inserting the unity: 
\begin{equation}
1=\int D\phi e^{-\int d^Dx \mathscr{L}_\phi} ,
\end{equation}
in the path-integral measure.%
\footnote{
It is shown that the effective field $\phi$ can be introduced without breaking the BRST symmetry.
In fact, it is shown \cite{Kondo01} that the operator $\mathcal{O}$ of mass dimension two is BRST-invariant up to the total derivative, i.e., $\bm{\delta} \mathcal{O}=\partial^{\mu} [ A_{\mu}^{a}(x) C^{a}(x) ]$ and  
 that the BRST transformation of $\phi$ is determined from the requirement $\bm{\delta} ( \phi+G\mathcal{O} )=0$.
}
See Fig.~\ref{fig:composite}. 
We observe the following:
\begin{itemize}
\item
 From the first term $\frac{1}{2} \phi^\dagger G^{-1} \phi$, we observe that $G$ represents the effective propagator of the \textbf{collective field} $\phi$, i.e., \textbf{ two-gluon  bound state} propagator. 

\item
The second term $\phi^\dagger \mathcal{O}$ yields the cubic interactions $\phi A A$ (and $\phi \Bar{C} C$) for the operator $\mathcal{O}$ quadratic in the off-diagonal gluons (and ghosts). 

\item
The third term $\frac{1}{2} \mathcal{O}^\dagger G \mathcal{O}$ involving only the fundamental fields has the form of an exchange of $\phi$ in the tree approximation. 
\end{itemize}

By including $\mathscr{L}_\phi$, the two-point functions $\Gamma_\Lambda^{(2)}$ are modified as
\begin{align}
\left( \Gamma_\Lambda^{(2)} \right)_{A_\mu^a A_\nu^b} &= W_\Lambda^\prime Q_{\mu \nu}^{ab} +\varphi \delta_{\mu \nu} \delta^{ab} , %\ \ (\varphi=\langle \phi \rangle) ,
\nonumber\\
\left(\Gamma_\Lambda^{(2)} \right)_{\Bar{C}^a C^b}  &=  -\Tilde{Z}_\Lambda \left( \mathscr{D}^2 \right)^{ab}+\alpha_\Lambda \varphi \delta^{ab} ,
\end{align}
where 
\begin{equation}
  \varphi=\langle \phi \rangle .
\end{equation}
Here we have adopted the truncation --- neglecting the four-point interactions 
among the off-diagonal gluons and off-diagonal ghosts.

\begin{figure}[t]
\begin{center}
\includegraphics[scale=0.22]{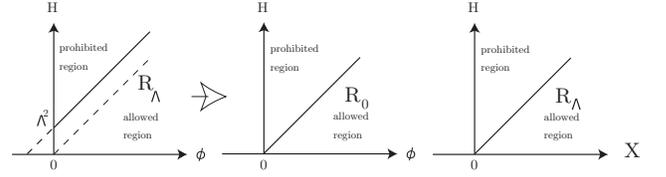}
\quad\quad
\end{center}
\vskip -0.5cm
\caption{ 
(Left panel)  The allowed region $\mathcal{R}_\Lambda$ and the prohibited region  in ($H, \varphi$) at $\Lambda > 0$   and $\Lambda=0$, 
(Right panel)  The allowed region $\mathcal{R}_\Lambda$ and the prohibited region in ($H, X$), where $X$ is equal to the shift of $\varphi$ by $-\Lambda^2$, 
$X:=\varphi+\Lambda^2$.
}
\label{fig:region} 
\end{figure}

We use the infrared regulator of the mass type and the same approximations for $W_\Lambda$, $\Tilde{Z}_\Lambda$ and $\alpha_\Lambda$ as those adopted in the previous case. Then we obtain the effective average potential $V_\Lambda (H, \varphi)$ describing the chromomagnetic   condensation and dynamical mass generation simultaneously.
We consider the simplest case of $\alpha_\Lambda \equiv 1$ to clarify  the qualitative feature (see \cite{Kondo03} for a physical meaning of the dimension-two condensate in this gauge).%
\footnote{
The thorough analysis including quantitative features will be given in a subsequent paper. 
}
In this case, the effective potential is given by
\begin{align}
V_\Lambda (H, \varphi)  =& \frac{1}{2g^2_\Lambda} H^2+ \frac{1}{2G_\Lambda} \varphi^2 +\Tilde{V}_\Lambda (H, \varphi) ,  
\end{align}
\begin{align}
\Tilde{V}_\Lambda (H, \varphi)
  =& -\frac{1}{4\pi^2} H^2 \left( \ln \frac{H}{\mu^2} -C \right) 
%\notag \\  &\times 
\Big[ \zeta \left( -1, \frac{3}{2}+\frac{X}{2H} \right)
\nonumber\\& 
+ \zeta \left( -1, -\frac{1}{2}+\frac{X}{2H} \right) \Big] \notag \\
 &+\frac{1}{4\pi^2} H^2 \Bigg[ \zeta^{(1,0)} \left( -1, \frac{3}{2}+\frac{X}{2H} \right) 
\nonumber\\& \quad\quad
+\zeta^{(1,0)} \left( -1, -\frac{1}{2}+\frac{X}{2H} \right) 
\nonumber\\& \quad\quad
 -2\zeta \left(-1, \frac{1}{2}+\frac{X}{2H} \right) \Bigg] , 
%\ X :=\varphi+\Lambda^2.
\end{align}
where 
\begin{equation}
 X :=\varphi+\Lambda^2  .
\end{equation}Here we have rescaled $H$ as $H \to \frac{1}{g} H$ for later convenience so that the quantum parts $\Tilde{V}_\Lambda$ does not include the $g$ dependence.
  We find that $\tilde{V}_\Lambda(H,\varphi)$ is obtained from $\tilde{V}_\Lambda(H)= \tilde{V}_\Lambda(H,\varphi=0)$ by shifting the variable $\Lambda^2 \to  \Lambda^2+\varphi$:
\begin{equation}
\tilde{V}_\Lambda(H,\varphi) = \tilde{V}_\Lambda(H,\varphi=0)|_{\Lambda^2 \to X }   =   \tilde{V}_\Lambda(H )|_{\Lambda^2 \to X } .
%\quad X := \varphi +\Lambda^2 .
\end{equation}               
The real-valuedness condition for $V_\Lambda$ is replaced by
\begin{equation}
  X-H  >0, \ \text{or}  \ \ H <  X   :=\varphi +\Lambda^2  .
\end{equation}
In other words, the stability excludes the region:
\begin{equation}
 H \ge     X:=\varphi +\Lambda^2  .
\end{equation}
Therefore, we define  the \textbf{allowed region for stability},
\begin{equation}
 \mathcal{R}_\Lambda = \left \{ (H, \varphi) ;  H <  X:=\varphi +\Lambda^2, H \ge0, \varphi> 0 \right \}.
\end{equation}
which is a region below the straight line $ H=  X$ with the slope $1$ and intercept $\Lambda^2$.
See Fig.~\ref{fig:region}.

%-------- Figure ----------------------------
%---------------------------------------------

%-------- Figure ----------------------------
\begin{figure}[ht]
\begin{center}
\includegraphics[scale=0.65]{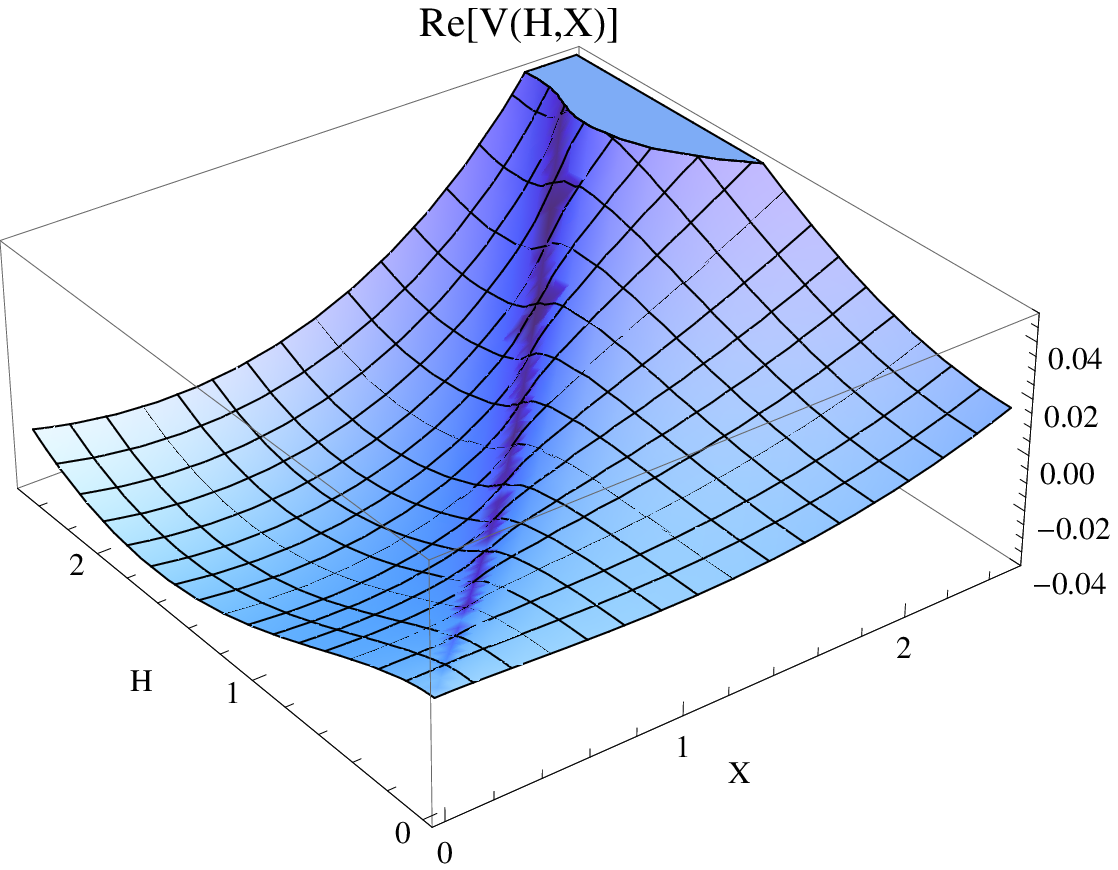}
\quad\quad
\includegraphics[scale=0.65]{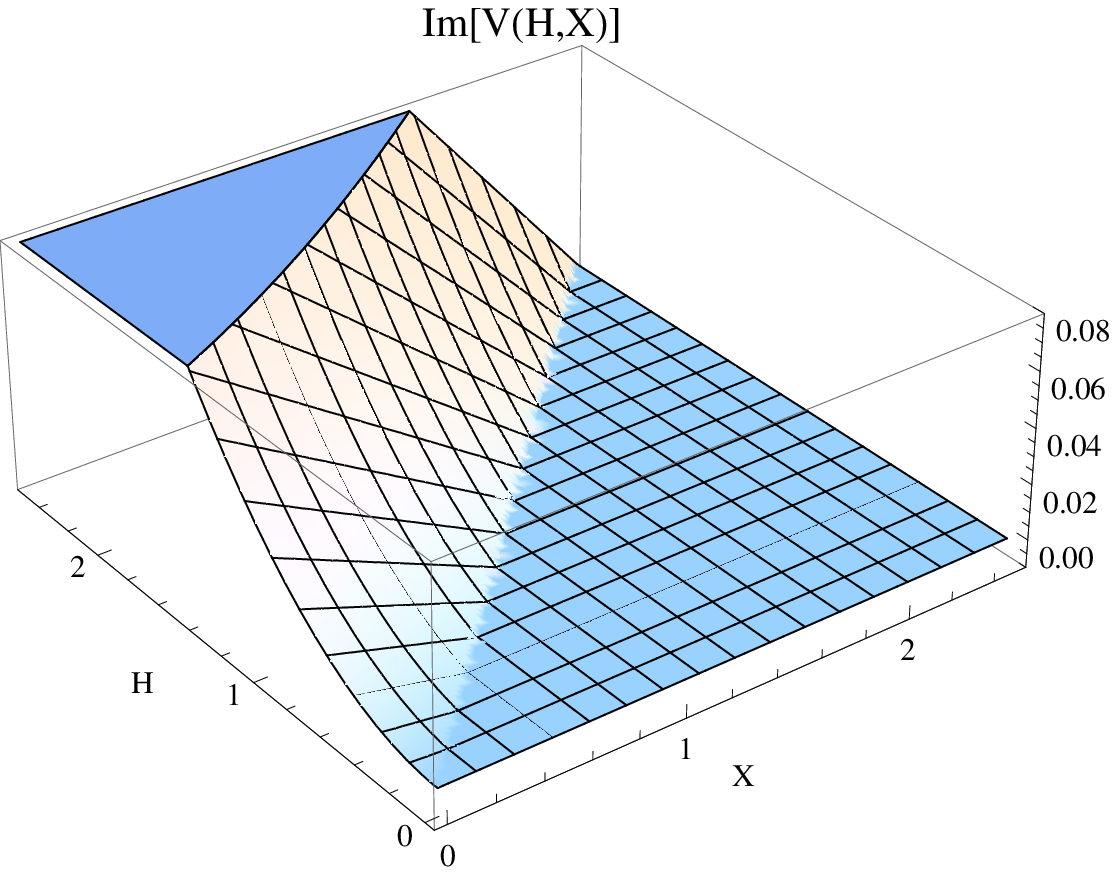}
\vskip 3mm
\includegraphics[scale=0.65]{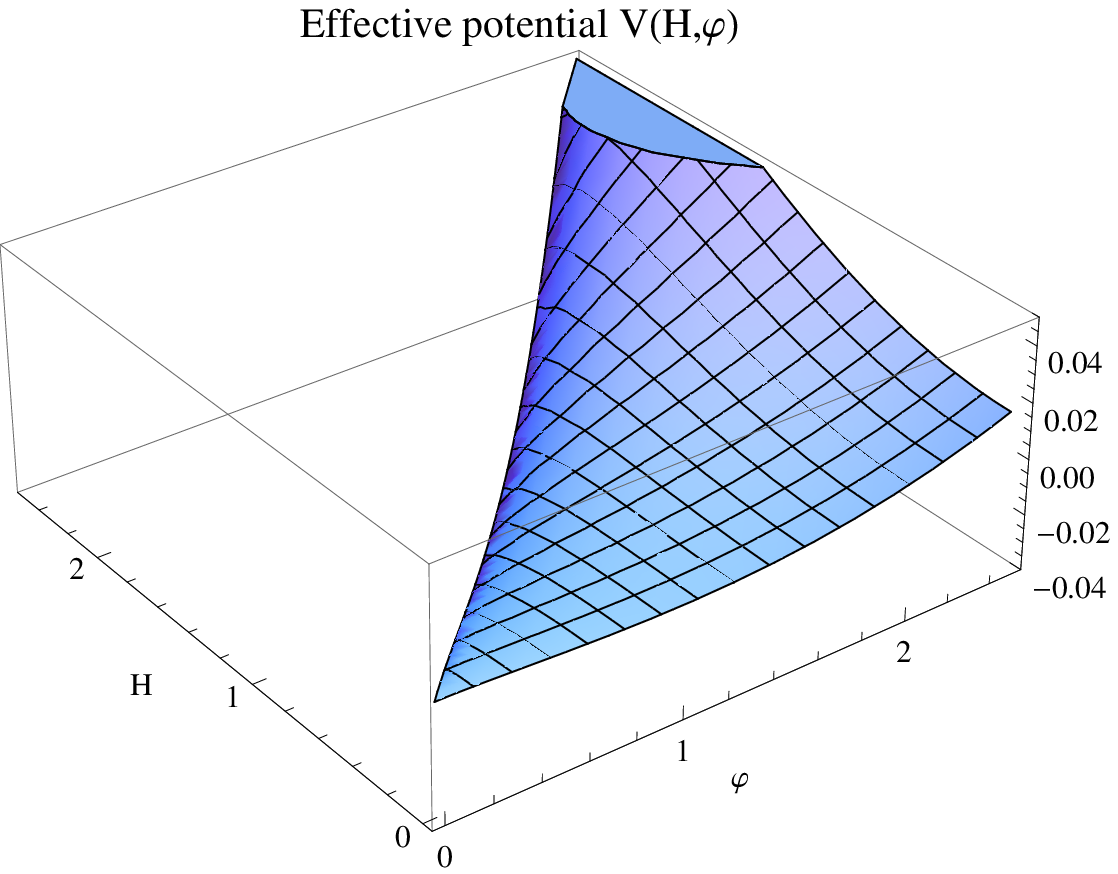}
\end{center}
\vskip -0.5cm
\caption{ 
The real  and imaginary parts of the quantum part of the effective potential $\tilde{V}(H,X)$.
The  region $0 \le H \le X$ is allowed where $\tilde{V}(H,X)$ is real valued, while the region  $X<H$ is prohibited where $\tilde{V}(H,X)$ includes the nonzero imaginary part. 
}
\label{fig:V(H,X)}
\end{figure}
%---------------------------------------------

$V_\Lambda (H,\varphi)$ can be made real valued by taking sufficiently large $\Lambda$, as in the case of $V_\Lambda (H)$.
In the absence of $\varphi$, this argument for eliminating the imaginary part does not work  in the  small $\Lambda$ region in which the inequality $ H>\Lambda^2$ is satisfied. This shortcoming is avoided by including $\varphi$. 
In fact,  the allowed region for stability $\mathcal{R}_\Lambda$  becomes narrower for a lower value of $\Lambda$, but survives even in the  limit $\Lambda \to 0$.
Hence, the $H$ axis  or $\varphi=0$ is excluded  in the  limit $\Lambda \to 0$.

The running coupling $g_\Lambda$ is monotonically increasing in decreasing $\Lambda$. Therefore,  the tree term $\frac{1}{2} g_\Lambda^{-2} H^2$ also becomes negligible for small enough $\Lambda$.

We can write down the flow equation for $G_\Lambda$.
Solving it, we find that $G^{-1}_\Lambda$   monotonically decreases  as $\Lambda$ decreases. Therefore, the effect of the tree term $\frac{1}{2} G_\Lambda^{-1} \varphi^2$ becomes more and more negligible for smaller $\Lambda$. 
In fact, the increasing of $G_\Lambda$ in decreasing $\Lambda$ is reasonable, since the bound state propagator $G_\Lambda (s)$ will approach the structure with a polelike dependence on $s$ for small enough $\Lambda$ \cite{Ellwanger94,Fukuda78,CS83}.
Therefore, the details of the behavior of $G_\Lambda$ does not change the following result qualitatively.

Thus the existence and location of the minimum can be dominantly determined by the quantum part $\Tilde{V} (H, \varphi)$. 
In view of these, we have looked  for the minimum of $\Tilde{V} _\Lambda (H, \varphi)$ in the region $\mathcal{R}_\Lambda$.
See Fig.~\ref{fig:V(H,X)} for the three-dimensional plot of  $\tilde{V}_\Lambda (H, X) $.
We find two minima: one minimum at 
$H \not= 0$ and $\varphi \not=0$ in the region $H>X$, 
and another minimum at 
$H=0$ and $\varphi \not=0$ in the region $H<X$. 
If we trust the above potential, the $H$ axis  or $\varphi=0$ is prohibited  in the  limit $\Lambda \to 0$, and therefore the former minimum is not allowed in the limit $\Lambda \to 0$, but it might be allowed by finding a more precise improved solution. 
The latter solution minimum survives in the limit $\Lambda \to 0$, which means that the mass generation occurs with the vanishing chromomagnetic condensation.

%\newpage
%-------- Figure ----------------------------
\begin{figure}[htbp]
\begin{center}
\includegraphics[scale=0.35]{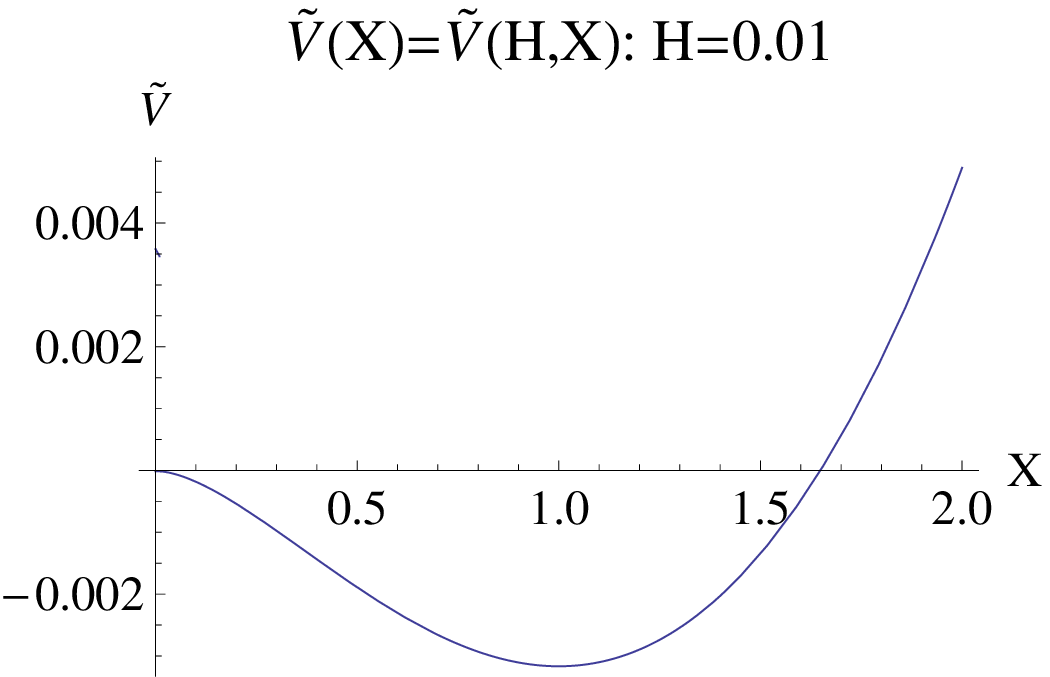}
\quad\quad
\includegraphics[scale=0.35]{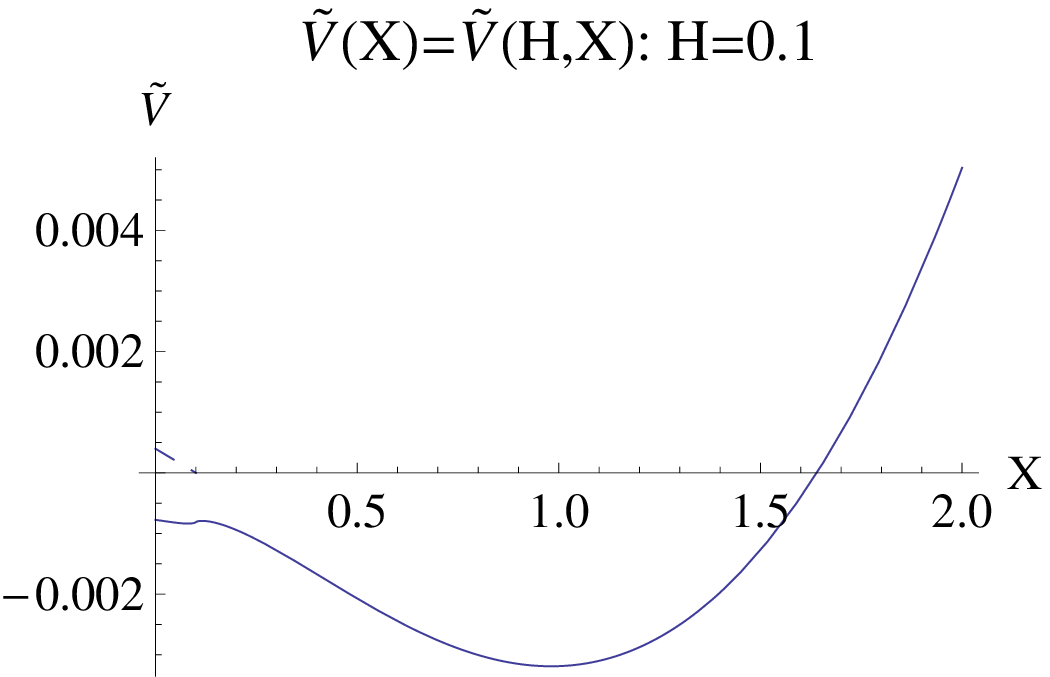}
\vskip 3mm
\includegraphics[scale=0.35]{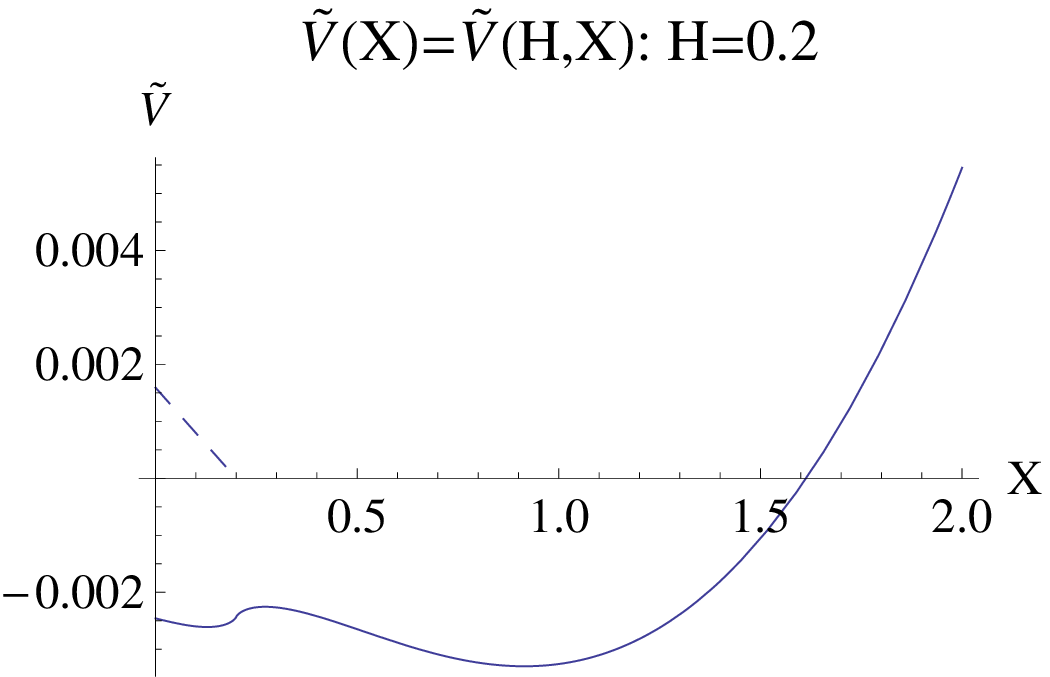}
\quad\quad
\includegraphics[scale=0.35]{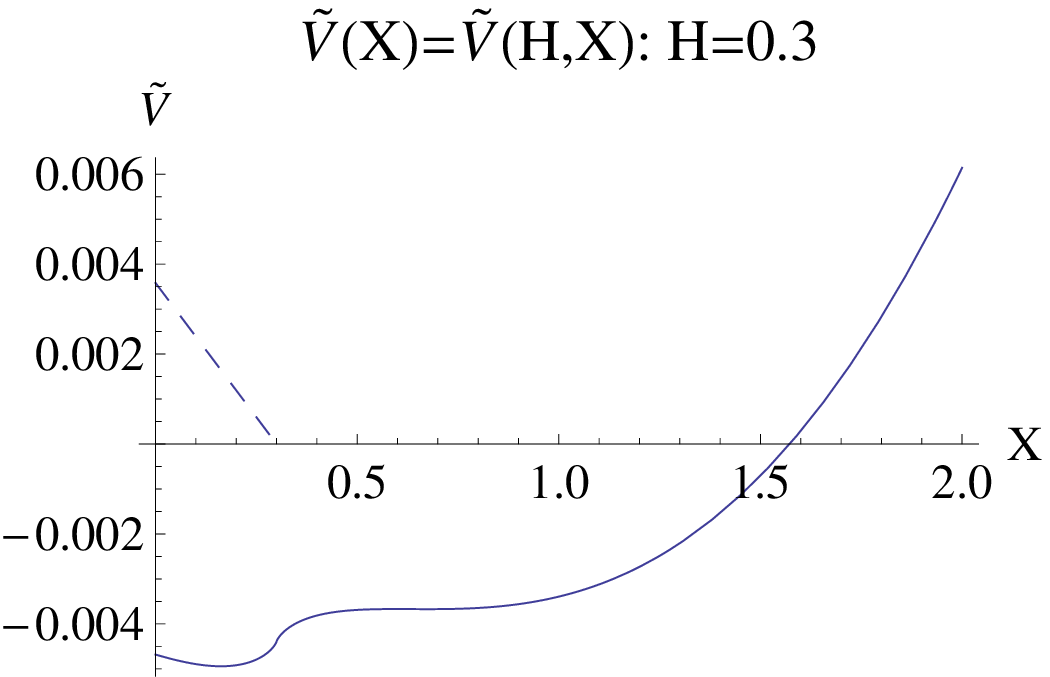}
\vskip 3mm
\includegraphics[scale=0.35]{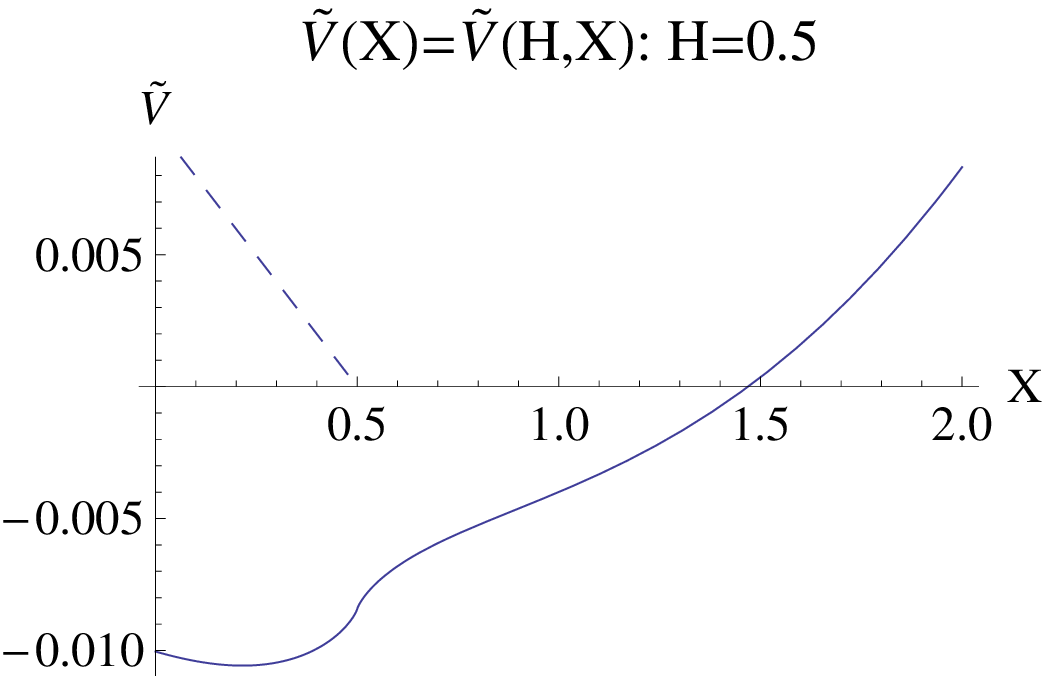}
\quad\quad
\includegraphics[scale=0.35]{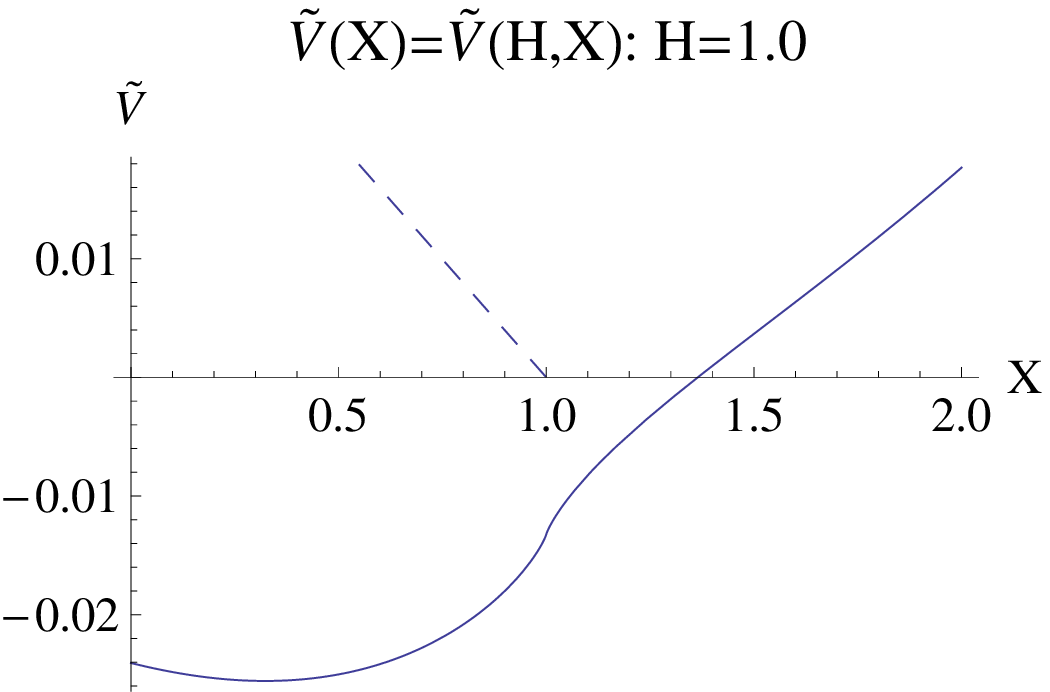}
%\includegraphics[scale=0.5]{fig-ep202/potential-q_H2.eps}
%\quad\quad
%\includegraphics[scale=0.5]{fig-ep202/potential-q_Hall.eps}
%\quad\quad
\end{center}
\vskip -0.5cm
\caption{ 
The real part (solid line) and imaginary part (dashed line) of the quantum part of the effective potential $\tilde{V}(X)=\tilde{V}(H,X)$ with various values of $H$:  
$H=0.01$, $H=0.1$, $H=0.2$, $H=0.3$, $H=0.5$, $H=1.0$.
%The  region $H  \le  X$ is allowed where $\tilde{V}(X)$ is real valued, while the region  $0<X<H$ is prohibited where $\tilde{V}(X)$ includes the nonzero imaginary part. 
}
\label{fig:V(X)}
\end{figure}
%---------------------------------------------

%-------- Figure ----------------------------
\begin{figure}[htbp]
\begin{center}
\includegraphics[scale=0.65]{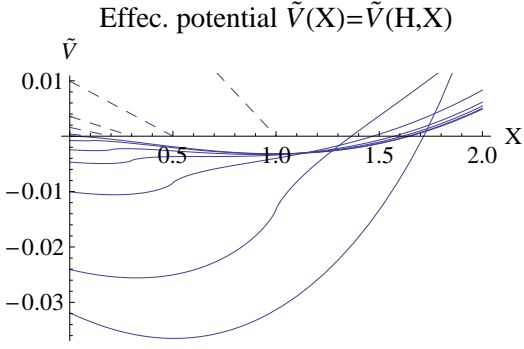}
\quad\quad
\end{center}
\vskip -0.5cm
\caption{ 
The real part (solid line) and imaginary part (dashed line) of the quantum part of the effective potential $\tilde{V}(X)=\tilde{V}(H,X)$ with various values of $H$:  
$H=0.01$, $H=0.1$, $H=0.2$, $H=0.3$, $H=0.5$, $H=1.0$.
The  region $H  \le  X$ is allowed where $\tilde{V}(X)=\tilde{V}(H,X)$ is real valued, while the region  $0<X<H$ is prohibited where $\tilde{V}(X)=\tilde{V}(H,X)$ includes the nonzero imaginary part. 
}
\label{fig:V(X)-all}
\end{figure}
%---------------------------------------------

%%%%%%%%%%%%%%%%%%%

%-------- Figure ----------------------------
\begin{figure}[htbp]
\begin{center}
%\includegraphics[scale=0.35]{fig-ep202/potential-q_0.eps}
%\quad\quad
\includegraphics[scale=0.35]{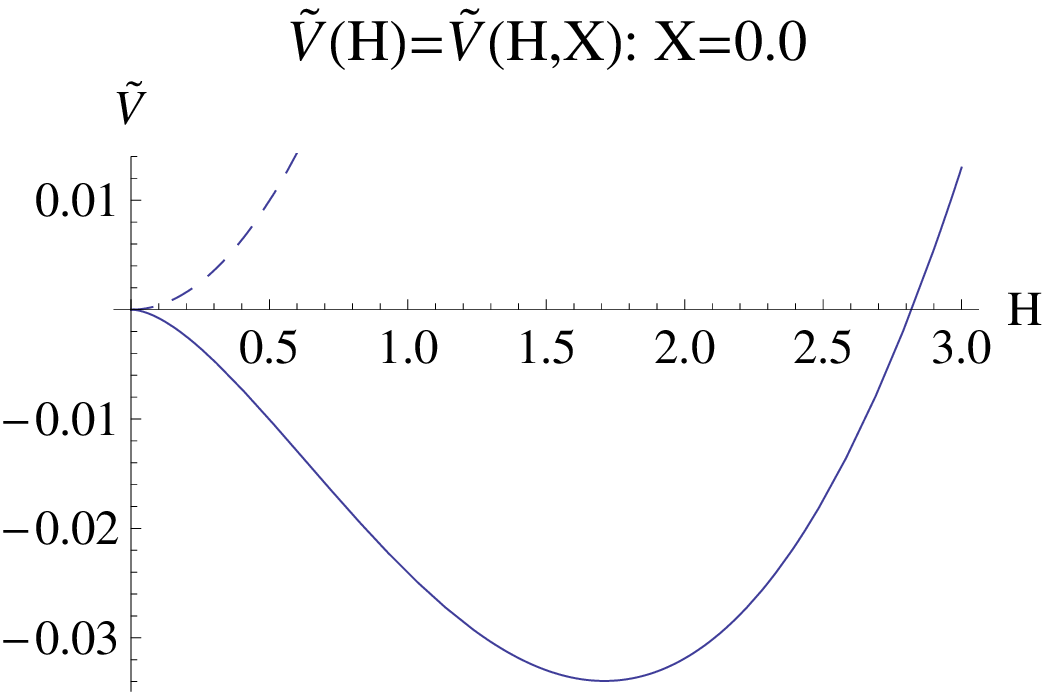}
\quad\quad
\includegraphics[scale=0.35]{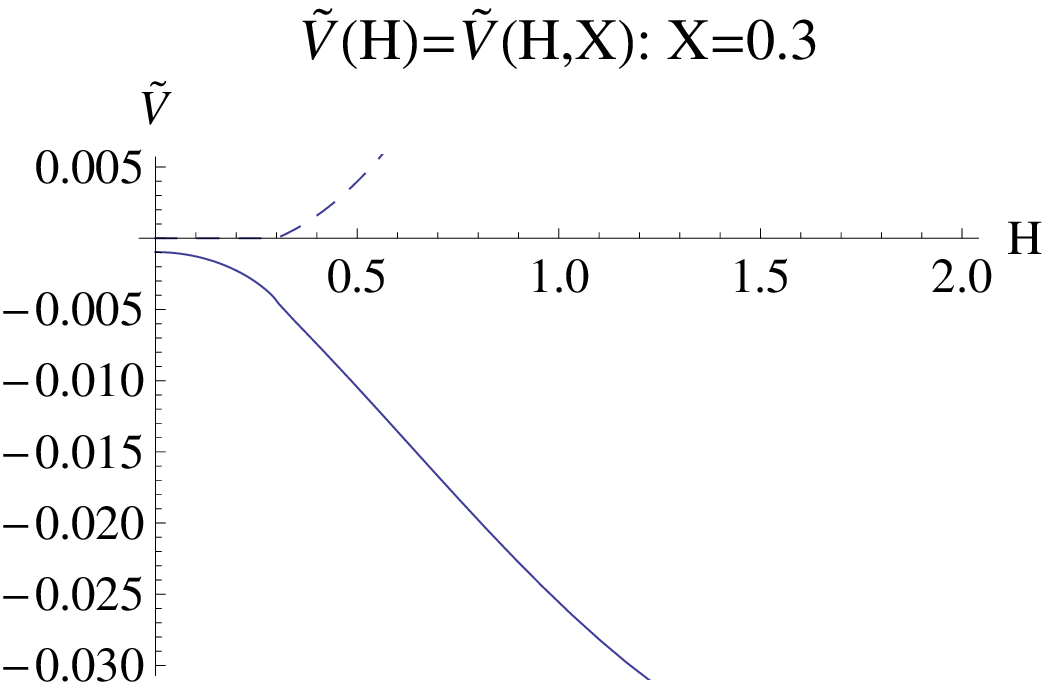}
\vskip 3mm
\includegraphics[scale=0.35]{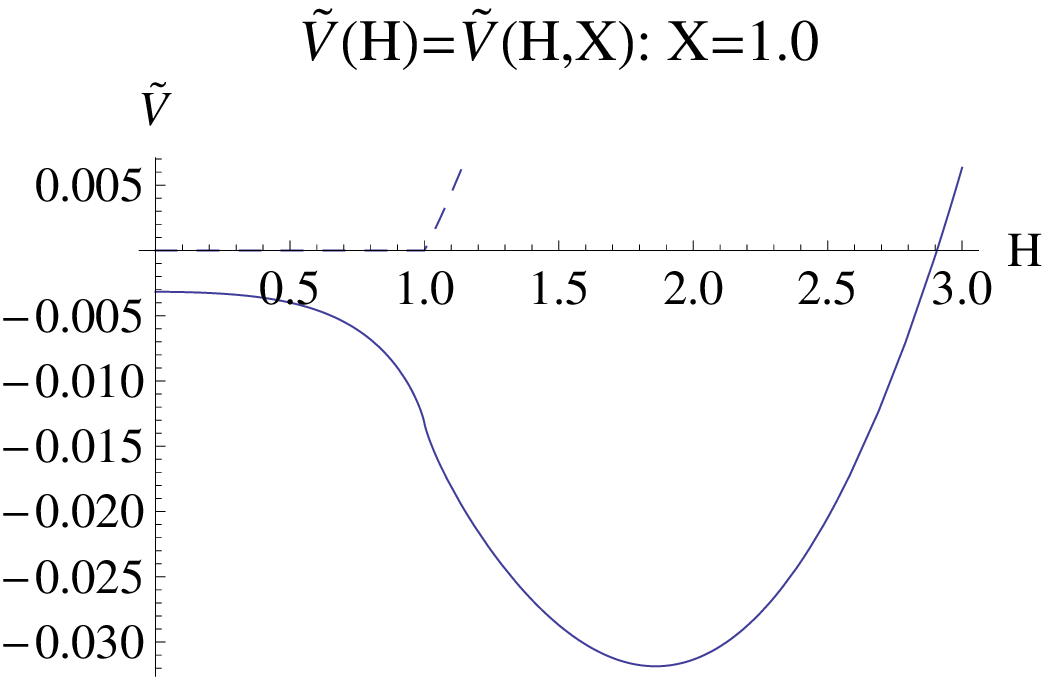}
\quad\quad
\includegraphics[scale=0.35]{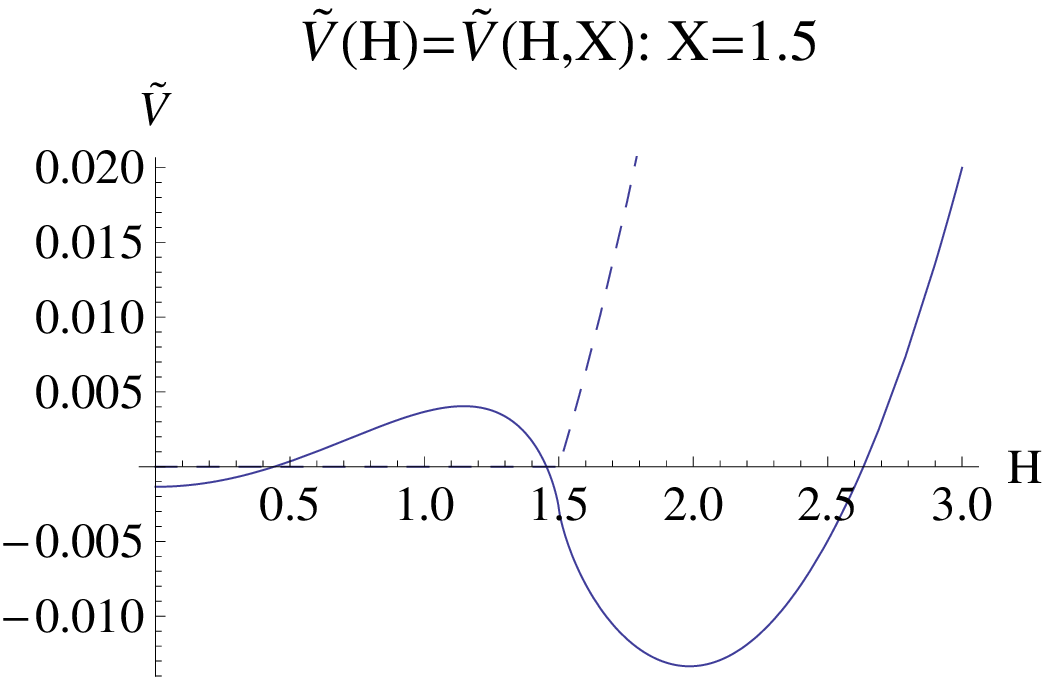}
\vskip 3mm
\includegraphics[scale=0.35]{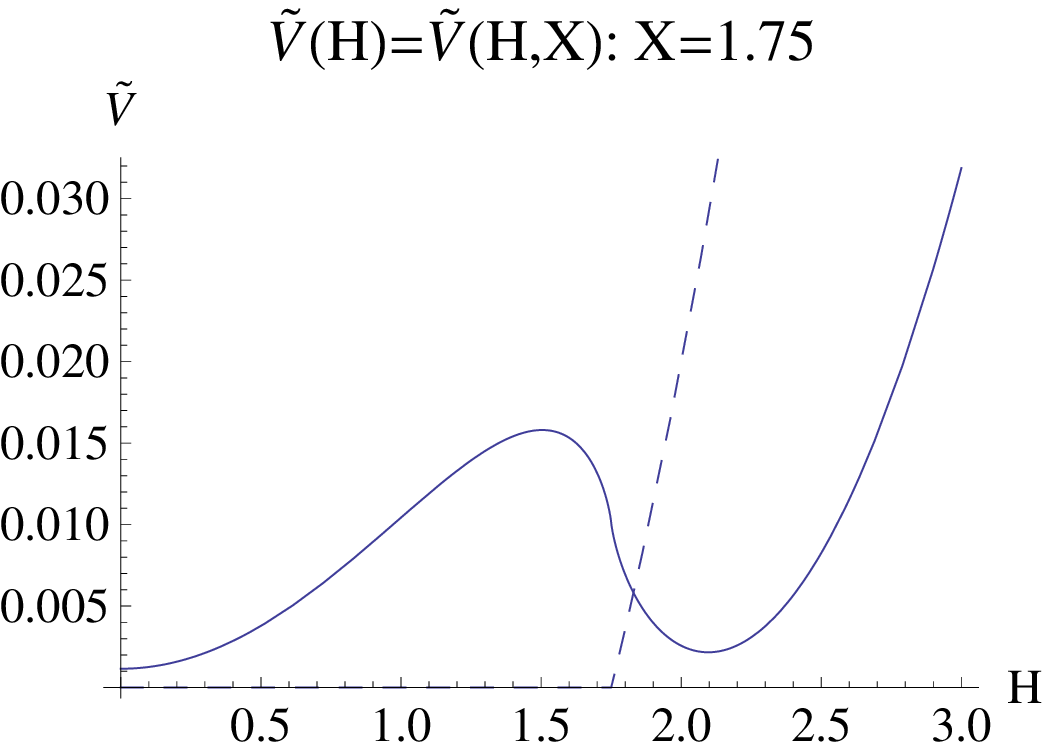}
\quad\quad
\includegraphics[scale=0.35]{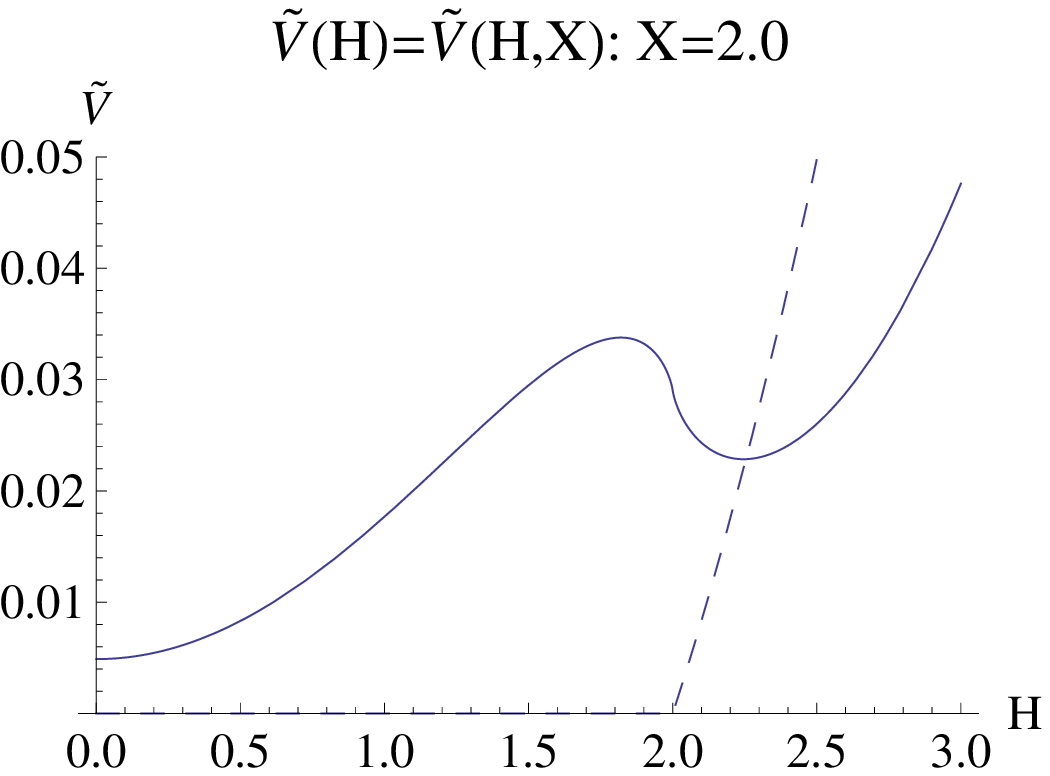}
%\includegraphics[scale=0.5]{fig-ep202/potential-q_H2.eps}
%\quad\quad
%\includegraphics[scale=0.5]{fig-ep202/potential-q_Hall.eps}
%\quad\quad
\end{center}
\vskip -0.5cm
\caption{ 
The real part  (solid line) and imaginary part (dashed line) of the quantum part of the effective potential $\tilde{V}(H)=\tilde{V}(H,X)$ with various values of $X$:  
$X=0$, $X=0.3$, $X=1.0$, $X=1.5$, $X=1.75$, $X=2.0$.
%The  region $0 \le H \le X$ is allowed where $\tilde{V}(H)$ is real valued, while the region  $X<H$ is prohibited where $\tilde{V}(H)$ includes the nonzero imaginary part. 
}
\label{fig:V(H)}
\end{figure}
%---------------------------------------------

%-------- Figure ----------------------------
\begin{figure}[htbp]
\begin{center}
\includegraphics[scale=0.65]{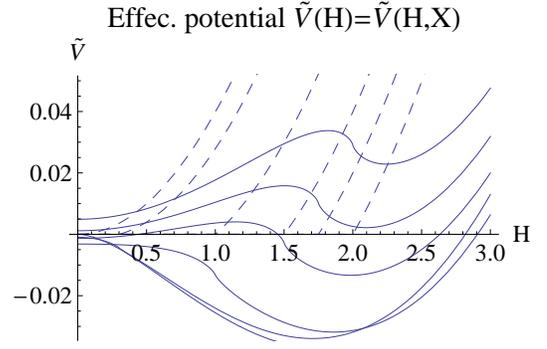}
\quad\quad
\end{center}
\vskip -0.5cm
\caption{ 
The real part (solid line) and imaginary part (dashed line) of the effective potential $\tilde{V}(H)=\tilde{V}(H,X)$ with various values of $X$:  
$X=0$, $X=0.3$, $X=1.0$, $X=1.5$, $X=1.75$, $X=2.0$.
The  region $0 \le H \le X$ is allowed where $\tilde{V}(H)$ is real valued, while the region  $X<H$ is prohibited where $\tilde{V}(H)$ includes the nonzero imaginary part. 
}
\label{fig:V(H)-all}
\end{figure}
%---------------------------------------------

%%%%%%%%%%%%%%%%%%%%%%%%%%

The following  are details of the potential. 
In Fig.~\ref{fig:V(X)}, we have given the plot of the potential  $\tilde{V}_\Lambda(X):=\tilde{V}_\Lambda(H,X)$ at fixed values of $H$.
The  region $X>H$ is allowed where $\tilde{V}(X)$ is real valued, while the region  $0<X<H$ is prohibited where $\tilde{V}(X)$ includes the nonzero imaginary part. 
For relatively small $H$,  a  lower (perturbative) minimum for the real part of the effective potential  exists for $X$ between zero and $H$, which is however in the prohibited region. 
This lower minimum is separated from the higher (nonperturbative) minimum by a little hill with a top at $X$ slightly above $H$. 
%The value of $X$ in the higher minimum decreases with increasing $H$. 
For higher $H$, a point is reached where the minimum with smaller $X$ has a lower energy than the one with greater $X$, around $H/\mu^2=0.2$.
%(We thus find a first-order phase transition around $H/\mu^2=0.2$.) 
For $H$ yet higher, the higher minimum disappears altogether and only the lower one remains. 
The full effective potential has the additional contribution $\frac{1}{2g^2}H^2 + \frac{1}{2G} \varphi^2$, which does not change the above picture, but rather strengthens the above tendency. 
Thus the minimum for the effective potential $V$ can exist for $H$ lower than a yet smaller critical value $H/\mu^2 \sim 0.3$.
See Fig.~\ref{fig:V(X)-all} for the collection of all the plots in  Fig.~\ref{fig:V(X)}. 

Thus, a nonzero chromomagnetic field decreases the effective gluon mass, and when the  chromomagnetic field is sufficiently strong
% a phase transition occurs, implying that 
the gluon mass disappears, since the lower minimum is in the prohibited region.
This conclusion is different from the result of Vercauteren and Verschelde \cite{VV08} that the mass is lowered to a value slightly lower than $gH$ after a phase transition occurred    when the  chromomagnetic field is strong enough.

%When considering the situation in which $H$ is an external field, applying such a field first lowers the value of the induced mass, and for $H$ above a certain value, the non-perturbative mass is destroyed, leaving only a perturbative value slightly smaller than $H$.
%However, this is not realized, since the effective action cannot have the imaginary part according to the analysis given above based on the functional renormalization group. 

In Fig.~\ref{fig:V(H)}, we have given the plot of the potential  $\tilde{V}_\Lambda(H):=\tilde{V}_\Lambda(H,X)$ at fixed values of $X$.
The  region $0<H<X$ is allowed where $\tilde{V}(H)$ is real valued, while the region  $H>X$ is prohibited where $\tilde{V}(H)$ includes the nonzero imaginary part. 
For relatively small $X$, a  lower (perturbative) minimum for the real part of the effective potential  exists for $H$ between zero and $X$, which is however in the prohibited region. 
When going to higher values of $X$, we find that $H=0$ turns into a local minimum of the potential.
For $H$ slightly below $X$, there is a maximum and for $H$ higher than $X$ there is a higher (non-perturbative) minimum.  
When increasing $X$, the  higher minimum first deepens out, reaching a lowest value for $X$, and it then goes up. 
%The value of $H$ in this point grows with increasing $X$.
%For $X$ large enough, this $H$ asymptotically goes to $X$. 
See Fig.~\ref{fig:V(H)-all} for the collection of all the plots in  Fig.~\ref{fig:V(H)}. 

We have considered the effect of the gluon mass on the chromomagnetic field condensation.
When the gluon mass is sufficiently large, the vacuum is no longer unstable against the formation of a homogeneous chromomagnetic field, and the Nielsen-Olesen instability, caused by the imaginary part in the effective potential,  is resolved. 
%But a sufficiently large gluon mass destroys the chromomagnetic field condensation.
%We conclude that switching a nonzero gluon mass first makes $H$ increase, and then destroy it completely.

%When considering both a homogeneous chromomagnetic field and a dimension-two condensation, the effective potential is minimized for zero chromomagnetic field $H=0$ with a non-perturbative value for the condensation $\varphi \ne 0$.  
%There are no unstable modes any longer, and the imaginary part in the potential is zero in this minimum.

\section{Conclusion and discussion}

In this paper, we have shown that the Nielsen-Olesen instability of the Savvidy vacuum with homogeneous chromomagnetic condensation is avoided in the framework of the FRG.
Actually, we have shown that the imaginary part of the effective average action vanishes  at sufficiently large infrared cutoff $\Lambda$, and this property can survive at $\Lambda=0$.
This behavior can be understood  as a fixed point solution of the flow equation for the complex-valued effective average action. 
Therefore, the  Nielsen-Olesen instability is an artifact of the  loop calculation in the perturbation theory.

First, the most important observation  given in Sect. II in this paper is the ``fixed point'' structure  that exists in the imaginary part $Im \Gamma_\Lambda$ of the complex-valued average effective action $\Gamma_\Lambda$ governed by the FRG equation of the Wetterich type.

This ``fixed point'' is different from the infrared   fixed point of the usual RG. 
%(I am afraid that the referee might misunderstand this point, as suggested from the comments 7) and 2)f).  I do not discuss the infrared fixed point in this paper at all.)
 The ``fixed point'' of this paper is restricted to the fixed point for all the scales from the ultraviolet down to the infrared, i.e., for any value of the flow parameter $\Lambda$, and the ``fixed point'' is considered only for the imaginary part of the complex-valued average effective action.  
In this sense, the claim of this paper is that \textbf{the complex valued FRG equation has the ``fixed point'' solution, i.e., the identically vanishing imaginary part $Im \Gamma_\Lambda \equiv 0$  as an exact solution}, while the real part $Re \Gamma_\Lambda$ does not have such a remarkable structure. This novel concept is schematically shown  in Fig.1 using the beta function $\beta(Im \Gamma_\Lambda)$ defined for the imaginary part $Im \Gamma_\Lambda$ of the average effective action $\Gamma_\Lambda$.

If the average effective action as the solution of FRG equation exhibits this fixed point structure, then the stability holds at any scale $\Lambda$ including $\Lambda=0$, since the imaginary part is identically vanishing and hence vanishing also at $\Lambda=0$.
This fact is the most important discovery of this paper, which has not been recognized in the preceding works to the best of the author's knowledge. 
In Fig.1, two possibilities for the solution are drawn: the fixed-point solution with $Im V \equiv 0$  and the nonfixed-point solution with $Im V \not\equiv 0$.

Second, we have proceeded to show that the solution of the FRG equation  satisfies the fixed point criterion. 
This is the content of Sect. IV. 
Of course, no one knows the exact solution of the FRG equation for the Yang-Mills theory. 
And we do not know even  the explicit analytical  form of the approximate  solution which is valid for any $\Lambda$. 
In order to examine the stability, however, it is enough  to show that all the solutions satisfy the fixed point structure at large but arbitrary value of  $\Lambda$ (for a finite interval of large $\Lambda$), since the smooth solutions must remain on the fixed point once they are on the ``fixed point,'' i.e., showing $Im V=0$ at large $\Lambda$.  See Fig.1.

For large $\Lambda$, in fact, we can find a good approximate analytical solution due to the asymptotic freedom of the theory, although this is not the case for small $\Lambda$ due to the strong interactions at the infrared region.  
Hence, the first check whether or not the solution satisfies the fixed point criterion, i.e., $Im V \equiv 0$ was performed in the large $\Lambda$ region  in Sect. IV. 
The result shows that the solution satisfies the criterion, i.e., no imaginary part at large but arbitrary value of  $\Lambda$ (for a finite interval of large $\Lambda$). 
This result is explicitly obtained for some infrared cutoff functions.  But it must hold for any infrared cutoff function on general ground, since the infrared cutoff function is required to satisfy the same asymptotic behavior for large $\Lambda$. 
Thus, the stability must be shown without discussing other details of the solution.
%hich is obtained up to the choice of the infrared cutoff function. 
The approximate solution given in this paper is valid for large $\Lambda$ at best $\Lambda > gH$. 
Therefore, Fig.2 is consistent with Fig.1.

To show the recovery of stability or the vanishing of the imaginary part just at $\Lambda=0$ starting from the  stability region at large $\Lambda$, we need to control the approximate solution along the flow from the large $\Lambda$ all the way down to $\Lambda=0$, which is quite a difficult task.
Fortunately, we do not need to do so for concluding only the stability.
%, since all such solutions must maintain the stability once they satisfy the stability for large $\Lambda$. 
%Of course, the best is to find the good approximate solution valid for any $\Lambda$.  

Next, we tried to find a better approximate  solution which is valid for even lower values of $\Lambda$ for understanding the physics behind the restoration of stability  or the elimination of instability.  
This is the content of Sect. V. 
%We have suggested that taking  account of the dynamical mass generation for (off-diagonal) gluons and ghosts can enlarge the stability region towards the smaller $\Lambda$.  
We have discussed the physical mechanism for keeping the stability for smaller $\Lambda$: the  stability is maintained even for small $\Lambda$ once  the mass generation occurs  for the off-diagonal gluons (and off-diagonal ghosts).

In fact, we have found two minima of the effective potential as a function of  the chromomagnetic field condensate $H$   and dynamical mass generation due to dimension-two vacuum condensation $\varphi$:

(i)   One minimum at 
$H \not= 0$ and $\varphi \not=0$ in the allowed region of stability $\mathcal{R}_\Lambda$ for  relatively small $\Lambda$:  
Both the chromomagnetic field condensate   and dynamical mass generation due to dimension-two vacuum condensation  occur  simultaneously in the region of validity for the infrared scale $\Lambda$.

(ii) Another minimum at 
$H=0$ and $\varphi \not=0$ in the region $\mathcal{R}_\Lambda$: 
This  minimum survives in  the limit $\Lambda \to 0$, which means that the mass generation occurs with the vanishing chromomagnetic condensation. 

If we accept our result for solving the flow equation at face value, however, our approximate solution for the effective action is valid only for the infrared cutoff $\Lambda$ above $\Lambda_0$, i.e.,  $\Lambda \ge \Lambda_0 \approx 0.335$ GeV.
In fact, the running Yang-Mills coupling constant $\alpha_\Lambda:=g_\Lambda^2/(4\pi)$ ceases  running at $\Lambda = \Lambda_0$ where $\alpha_\Lambda=\alpha^0 \approx 1.88...$. 
This is the same situation as that encountered in the work \cite{RW97}.

In order to obtain the true effective action, we need to solve the flow equation all the way down to $\Lambda \to 0$.
In fact,  a finite value for  the running coupling constant has been obtained even at $\Lambda=0$   in the framework of the FRG \cite{Gies02,EGP11}, although it had been shown for the first time  in the framework of the Schwinger-Dyson equation  \cite{scaling}.

The comparison of our result for the effective potential with that of \cite{EGP11}  suggests that  (i)  $H \not= 0$ and $\varphi \not=0$ is realized in the Yang-Mills vacuum. 
Using these solutions \cite{Gies02,EGP11}, moreover, we are able to discuss the possible relationship between the stability and  the scaling/decoupling \cite{{scaling},decoupling,decoupling-lattice} solutions which are recently claimed to be  the true infrared solutions in the deep infrared region realizing quark and gluon confinement \cite{FMP09,BGP10,FP13,Kondo11}.
These issues will be further discussed in future works. 

{\it Acknowledgements}\ ---
The author would like to thank Holger Gies for private communications on the works \cite{Gies02,RW94,RW97} 
and kind hospitality from him in the stay of Jena (September 2013). 
He thanks also Jan Pawlowski for giving  constructive  comments on the issues discussed in this paper and kind hospitality to him in the stay of Heidelberg (September 2013). 
Thanks are also due to Daniel  Litim for giving valuable comments on   some technical issues  and kind hospitality to him in the stay of Sussex University at Brighton (September 2013).
This work is  supported by Grant-in-Aid for Scientific Research (C)  24540252 from Japan Society for the Promotion of Science (JSPS).

\appendix
\section{Decomposition of a complex-valued matrix}

In order to obtain the inverse matrix $P + i Q$ of the complex matrix $A + i B$, we set
\begin{equation}
 (A + i B) (P + i Q) = \bm{1} = ( P + i Q ) (A + i B) , 
\end{equation}
which yields
\begin{align}
\left\{ \begin{array}{l}
AP - BQ = \bm{1} = PA - QB ,\\
AQ + BP = 0 =  PB + QA .
\end{array} \right. 
\end{align}
From the second equation, we obtain 
\begin{equation}
Q = - A^{-1} BP = - PB A^{-1} .
\end{equation}
Substituting this relation into the first equation to eliminate $Q$, we obtain $P$:
\begin{align}
  &  AP + B A^{-1} BP =  \bm{1} = PA + PB A^{-1} B  
\nonumber\\
 \Rightarrow  & (A  + B A^{-1} B)P =  \bm{1} = P(A + B A^{-1} B)  
\nonumber\\
 \Rightarrow  & P =   (A + B A^{-1} B)^{-1} 
%= A ( (A+ B A^{-1} B) A )^{-1} 
\nonumber\\  &
=  A (  AA+ B A^{-1} BA )^{-1} 
=   (  AA+ AB A^{-1} B  )^{-1} A  ,
\end{align}
and hence
\begin{align}
Q =& - (AB^{-1}A+B)^{-1}
\nonumber\\  
=& - A^{-1} B A ( AA + B A^{-1} BA )^{-1} 
\nonumber\\  
=&   -  ( AA + AB A^{-1} B )^{-1} A B A^{-1}
%= - (A + B A^{-1} B)^{-1} B A^{-1} 
 .
\end{align}
If $[A , B] = 0$, i.e., $AB=BA$, then  $B^{-1}A^{-1}=A^{-1}B^{-1}$, which leads to $A^{-1}B=BA^{-1}$ and $AB^{-1}=B^{-1}A$.
Therefore, we obtain
\begin{equation}
 P =  ( A^{2} + B^{2} )^{-1} A, \quad
 Q = - (A^{2} + B^{2} )^{-1} B 
.
\end{equation}
Note that if $B \to 0$, then $P \to A^{-1}$ and $Q \to 0$.

\section{Removing the ultraviolet divergence}

The ultraviolet divergence of (\ref{flow-eq1}) at $\tau=0$  is removed as follows. 
 (i) We introduce the parameter: %$\epsilon:= 2-\frac{D}{2}$ 
\begin{equation}
\epsilon := 2-\frac{D}{2} = \frac{4-D}{2} \quad  ( D= 4-2\epsilon) ,
\end{equation}
and replace $D$ by $D=4-2\epsilon$. 
(ii) Expand the right-hand side into the Laurent series in powers of $\epsilon$, and  (iii) extract the terms of order $\epsilon^0$ ($\epsilon$-independent terms).

By using the rescaling of $\tau$,  the flow equation (\ref{flow-eq1}) reads
\begin{align}
   \partial_{t} \Gamma_{\Lambda}  
 =&  \frac{N}{2} \frac{(2gH)^{\frac{D}{2}-1}}{(4\pi)^{\frac{D}{2}}}  \Big\{  ( W_{\Lambda}^{\prime} )^{-1}  ( 2 - \eta_{\Lambda} ) Z_{\Lambda} \Lambda^{2}  
\nonumber\\&  
\times  \int_{0}^{\infty} d s \ s^{1 - \frac{D}{2}} e^{- s \frac{Z_{\Lambda} \Lambda^{2}}{2 W_{\Lambda}^{\prime}gH}}  
\nonumber\\&  
\frac{ (D-3)e^{-\frac12 s}+e^{-\frac32 s }+e^{\frac12 s }}{1-e^{-s}}   
%\nonumber\\&
%-   ( W_{\Lambda}^{\prime} )^{1- \frac{D}{2} }  ( 2 - \eta_{\Lambda} ) Z_{\Lambda}  \Lambda^{2}  \int_{0}^{\infty} d \tau \tau^{1 - \frac{D}{2}}  e^{- \tau Z_{\Lambda} \Lambda^{2}}  \frac{  e^{-\tau W_{\Lambda}^{\prime} gH} }{1-e^{-2\tau W_{\Lambda}^{\prime} gH}} 
\nonumber\\&
+   \alpha_\Lambda   ( 2 - \eta_{\Lambda} ) Z_{\Lambda}  \Lambda^{2}  
\nonumber\\&  
\times \int_{0}^{\infty} d s \  s^{1 - \frac{D}{2}}  e^{- s \frac{Z_{\Lambda} \Lambda^{2}}{2\alpha_\Lambda^{-1} gH} }  \frac{  e^{-\frac12 s} }{1-e^{-s}} 
\nonumber\\&
 - (\tilde{Z}_{\Lambda})^{-1} (2 - \tilde{\eta}_\Lambda) \tilde{Z}_{\Lambda} \Lambda^{2} 
\nonumber\\&  
\times \int_{0}^{\infty} d s \ s^{1 - \frac{D}{2}}  e^{- s \frac{\tilde{Z}_{\Lambda} \Lambda^{2}}{2gH} }  \frac{2e^{-\frac12 s }}{1-e^{-s}}  \Big\} .
\label{flow-eq2}
\end{align}

We introduce the \textbf{generalized Riemann $\zeta$-function} or the \textbf{Hurwitz $\zeta$-function} $\zeta (z, \lambda)$ defined by 
\begin{equation}
\zeta (z, \lambda) :=\sum \limits_{n=0}^{\infty} \frac{1}{(n+\lambda)^z} ,
\label{zeta-def}
\end{equation}
which has its integral representation \cite{EORBZ94}: 
\begin{equation}
\zeta (z, \lambda) = \frac{1}{\Gamma(z)} \int^\infty_0 ds \ s^{z-1} \frac{e^{-\lambda s}}{1-e^{-s}} \ \ ( \text{Re} z>1 , \text{Re} \lambda>0 ) .
\label{integral-rep-zeta}
\end{equation}
Although the Hurwitz $\zeta$-function $\zeta (z , \lambda)$ is originally defined for ${\rm Re} \ z > 1 , \ {\rm Re} \ \lambda > 0$, 
it can be analytically continued to other region in the complex $z$-plane as an analytic function.
Then the flow equation is rewritten as
\begin{align}
\partial_{t} \Gamma_{\Lambda}  &=  \frac{N}{2} \frac{(2gH)^{\frac{D}{2}-1}}{(4\pi)^{\frac{D}{2}}} 
\Gamma \left( 2-\frac{D}{2} \right) 
\Biggr\{  ( W_{\Lambda}^{\prime} )^{-1}  ( 2 - \eta_{\Lambda} ) Z_{\Lambda} \Lambda^{2}  
\nonumber\\&  
\times  %\nonumber\\ & \times
    \Bigg[ ( D-3 )  \zeta \left( 2-\frac{D}{2}, \frac{1}{2} +\frac{Z_{\Lambda} \Lambda^{2}}{2 W_{\Lambda}^{\prime}gH}  \right) \nonumber\\ 
                        &+\zeta \left( 2-\frac{D}{2}, \frac{3}{2}+\frac{Z_{\Lambda} \Lambda^{2}}{2 W_{\Lambda}^{\prime}gH}\right)
\nonumber\\&  +\zeta \left( 2-\frac{D}{2}, -\frac{1}{2}+\frac{Z_{\Lambda} \Lambda^{2}}{2 W_{\Lambda}^{\prime}gH} \right) \Bigg]   
\nonumber\\&
+   \alpha_\Lambda   ( 2 - \eta_{\Lambda} ) Z_{\Lambda}  \Lambda^{2}  \zeta \left( 2-\frac{D}{2}, \frac{1}{2} +\frac{Z_{\Lambda} \Lambda^{2}}{2\alpha_\Lambda^{-1} gH} \right) 
\nonumber\\&
 -  2 (\tilde{Z}_{\Lambda})^{-1} (2 - \tilde{\eta}_\Lambda) \tilde{Z}_{\Lambda} \Lambda^{2} \zeta \left( 2-\frac{D}{2}, \frac{1}{2} +\frac{ \Lambda^{2}}{2 gH} \right)  \Biggr\} .
\label{flow-eq3}
\end{align}
This expression has the divergence at $D=4$,
since $\Gamma (0) =\infty$, although $\zeta (0, \lambda) <\infty  $ for $\lambda  <\infty  $.

Therefore, we  rewrite the flow equation into 
\begin{align}
\partial_{t} \Gamma_{\Lambda}   =&  \frac{N}{2} \frac{(2gH)^{1-\epsilon}}{(4\pi)^{2-\epsilon}} 
\Gamma \left( \epsilon \right) 
\Biggr\{  ( W_{\Lambda}^{\prime} )^{-1}  ( 2 - \eta_{\Lambda} ) Z_{\Lambda} \Lambda^{2}  
\nonumber\\&  
\times  %\nonumber\\ & \times
    \Bigg[ ( 1-2\epsilon )  \zeta \left( \epsilon, \frac{1}{2} +\frac{Z_{\Lambda} \Lambda^{2}}{2 W_{\Lambda}^{\prime}gH}  \right) \nonumber\\ 
                        &+\zeta \left( \epsilon, \frac{3}{2}+\frac{Z_{\Lambda} \Lambda^{2}}{2 W_{\Lambda}^{\prime}gH}\right)+\zeta \left( \epsilon, -\frac{1}{2}+\frac{Z_{\Lambda} \Lambda^{2}}{2 W_{\Lambda}^{\prime}gH} \right) \Bigg]   
\nonumber\\&
+   \alpha_\Lambda   ( 2 - \eta_{\Lambda} ) Z_{\Lambda}  \Lambda^{2}  \zeta \left( \epsilon, \frac{1}{2} +\frac{Z_{\Lambda} \Lambda^{2}}{2\alpha_\Lambda^{-1} gH} \right) 
\nonumber\\&
 - 2 (\tilde{Z}_{\Lambda})^{-1} (2 - \tilde{\eta}_\Lambda) \tilde{Z}_{\Lambda} \Lambda^{2} \zeta \left( \epsilon, \frac{1}{2} +\frac{ \Lambda^{2}}{2 gH} \right)  \Biggr\} .
\label{flow-eq4}
\end{align}

 For $\epsilon \ll 1 $, we can use the expansions around $\epsilon=0$:
\begin{align}
\Gamma ( \epsilon) =&  \epsilon^{-1} - \gamma   + O (\epsilon), \\
\mu^{2\epsilon} \left( \frac{2gH}{4\pi} \right)^{-\epsilon}  =& \exp \left[ -\epsilon \ln \left(\frac{2gH}{4\pi \mu^2} \right) \right] 
\nonumber\\   
=& 1-\epsilon \ln \frac{2gH}{4\pi \mu^2}+O(\epsilon^2) ,
\end{align}
and
\begin{equation}
\zeta( \epsilon , \lambda) = \zeta(0, \lambda) + \epsilon \ \zeta^{(1,0)}(0, \lambda) + O(\epsilon^2) ,
\end{equation}
where we have defined 
\begin{equation}
\zeta^{(m,n)} (z, \lambda):= \frac{\partial^m}{\partial z^m} \frac{\partial^n}{\partial \lambda^n}  \zeta (z, \lambda) .
\label{zeta-def-derivative}
\end{equation}

\section{Generalized Riemann $\zeta$-function}

The generalized Riemann $\zeta$-function $\zeta^{(0,0)} (1-n, \lambda)$ can be represented as
\begin{align}
%  \zeta^{(0,0)} (-m, \lambda) =  - \frac{1}{m+1}  B_{m+1}(\lambda)   ,
%\ 
  \zeta^{(0,0)} (1-n, \lambda) =  - \frac{1}{n}  B_{n}(\lambda)  ,
\end{align}
where $B_{n}(\lambda)$ is the Bernoulli polynomial of degree $n$. 

For $n=1$, 
\begin{align}
  \zeta^{(0,0)} (0, \lambda) =  -  B_{1}(\lambda)  = - \lambda +  \frac{1}{2} .
\end{align}
For $n=2$, 
\begin{align}
  \zeta^{(0,0)} (-1, \lambda) =  - \frac{1}{2}  B_{2}(\lambda)  =  - \frac{1}{2}  \left( \lambda^2 - \lambda +\frac{1}{6} \right)  .
\end{align}

The expansion of  the derivative of the generalized Riemann $\zeta$-function $\zeta^{(1,0)} (1-n, \lambda)$  for large $\lambda$ is given by
\cite{EORBZ94}
% [Elizalde et al., 1994] 12--13 pages,
\begin{align}
  &
  \zeta^{(1,0)} (1-n, \lambda) \nonumber\\  
=& \frac{1}{n} \left( \ln \lambda - \frac{1}{n}  \right) B_n(\lambda) - \frac{1}{2n} \lambda^{n-1} 
\nonumber\\ &
- \frac{1}{n} \sum_{k=2}^{n} B_k \sum_{j=0}^{k-1} {}_n{}C_j \frac{(-1)^j}{k-j} \lambda^{n-k} 
\nonumber\\ &
+ (-1)^{n-1} (n-1)! \sum_{k=n+1}^{\infty} \frac{B_k}{k(k-1) \dots (k-n)} \lambda^{n-k} ,
\end{align}
where $B_k$ is the Bernoulli numbers. 

For $n=1$, 
\begin{align}
&  \zeta^{(1,0)} (0, \lambda) 
\nonumber\\  
=&   \left( \ln \lambda -1  \right) B_1(\lambda) - \frac{1}{2}  
+ \frac{1}{2}  B_2 \lambda^{-1}  + O(\lambda^{-2})
\nonumber\\  
=&   \left( \ln \lambda -1  \right) \left(   \lambda -  \frac{1}{2}  \right) - \frac{1}{2}  + O(\lambda^{-1})
 .
\end{align}
For $n=2$, 
\begin{align}
&  \zeta^{(1,0)} (-1, \lambda) 
\nonumber\\  
=&   \frac{1}{2}  \left( \ln \lambda - \frac{1}{2}  \right) B_2(\lambda) - \frac{1}{4}  \lambda
+ \frac{1}{2}  B_2 \lambda^{-1}   
%\nonumber\\  &
+ O(\lambda^{-2})
\nonumber\\  
=&    \frac{1}{2}  \left( \ln \lambda - \frac{1}{2}  \right)   \left( \lambda^2 - \lambda +\frac{1}{6} \right)  - \frac{1}{4}  \lambda
%\nonumber\\ &
+   O(\lambda^{-1})
 .
\end{align}

The following recursion relation holds\cite{EORBZ94}:
\begin{equation}
\zeta^{(1,0)} (-1, a+n)=\zeta^{(1,0)} (-1, a)+ \sum \limits_{n=0}^{n-1} (k+a) \ln (k+a)  .
\end{equation}
In particular, 
\begin{equation}
\zeta^{(1,0)} (-1, a+1) = \zeta^{(1,0)} (-1, a) + a \ln a .
\end{equation}

\section{Flow of   gauge parameters in   MA gauge}

It was shown \cite{Schaden99,SIK01,KS01,EW02} that the gauge-fixing parameter $\beta$ of the diagonal part in the Lorentz gauge obeys the RG equation to the one-loop calculation:  
\begin{equation}
\mu\frac{\partial}{\partial\mu} \beta_{\rm R}
 =\frac{44}3 \beta_{\rm R}\frac{g_{\rm R}^2}{(4\pi)^2} ,
\label{MAbeta}
\end{equation}
and that the gauge-fixing parameter $\alpha$ of the off-diagonal part in the modified maximal Abelian (MA) gauge obeys the RG equation: 
\begin{equation}
\mu\frac{\partial}{\partial\mu} \alpha_{\rm R}
 =
  \left[-2\alpha_{\rm R}^2+\frac83\alpha_{\rm R}-6
  \right] \frac{g_{\rm R}^2}{(4\pi)^2} .
\label{MAalpha}
\end{equation}

It is well known that the running of the gauge coupling constant is governed by the differential equation:
\begin{align}
   \beta(g_{\rm R}^2) := \mu {\partial g_{\rm R}^2 \over \partial \mu} 
 =-  {22 \over 3}\frac{C_2(G)}{(4\pi)^2}  g_{\rm R}^4  ,
\label{betaf2}
\end{align}
Equation (\ref{betaf2}) is a closed equation for $g_R$, which is solved exactly as a function of $\mu$: 
\begin{equation}
  g_R^2(\mu) = {g_R^2(\mu_0) \over 1+ {22 \over 3}{C_2(G) \over (4\pi)^2}g_R^2(\mu_0) \ln {\mu \over \mu_0}} 
  = {1 \over {22 \over 3}{C_2(G) \over (4\pi)^2} \ln {\mu \over \Lambda_{\rm QCD}}}  ,
\label{betasol}
\end{equation}
where we have used the boundary condition $g_R(\mu_0)=\infty$ at $\mu_0=\Lambda_{\rm QCD}$. 
 Using the solution (\ref{betasol}), the derivative ${1 \over g_R^2} \mu {\partial \over \partial \mu}$ in (\ref{MAbeta}) and (\ref{MAalpha})  is rewritten as
\begin{equation}
  {1 \over g_R^2} \mu {\partial \over \partial \mu}
% = {22 \over 3}{C_2(G) \over (4\pi)^2} \ln {\mu \over \Lambda_{\rm QCD}}  \mu {\partial \over \partial \mu}
  =  {22 \over 3}{C_2(G) \over (4\pi)^2}   {\partial \over \partial \ln \ln {\mu \over \Lambda_{\rm QCD}}} .
\label{difftransf}
\end{equation}

%%%%%%%%%%%%%%%%%%%%%%%%%%%%%%%%%%%%%%%%%%%%%%%%%%%%%%%%%%%%

\begin{figure}[tbp]
\begin{center}
\includegraphics[scale=0.3]{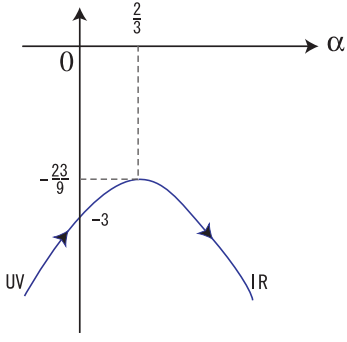}
\caption{
%(left panel) 
The flow of the gauge parameter $\alpha$ in the modified MA gauge for the off-diagonal gauge field.  
%(right panel) The flow of the gauge parameter $\beta$ in the Lorentz gauge for the diagonal field.  
The arrow is directed to the IR region, i.e., decreasing the renormalization scale $\mu$.
}
\label{fig:GaugeParameterFlowMA}
\end{center}
\end{figure}

%%%%%%%%%%%%%%%%%%%%%%%%%%%%%%%%%%%%%%%%%%%%%%%%%%%%%%%%%%%%

We apply  (\ref{difftransf}) to rewrite the differential equation (\ref{MAbeta}) into the form which does not explicitly depend on $g^2$: \begin{align}
{22 \over 3} 2   {\partial \over \partial \ln \ln {\mu \over \Lambda_{\rm QCD}}}  \beta_{\rm R}
 =\frac{44}3  \beta_{\rm R} ,
\end{align}
which is easily solved.
 The integration,
\begin{align}
 \int_{\beta}^{\bar{\beta}} {d \beta \over \beta} = \int_{\mu_0}^{\mu} d \ln \ln {\mu \over \Lambda_{\rm QCD}} ,
\end{align}
yields 
\begin{align}
  \bar{\beta} =  \beta {\ln (\mu/\Lambda_{\rm QCD})  \over \ln (\mu_0/\Lambda_{\rm QCD}) } 
%=  \beta  \left(  {\bar{g}^2 \over g^2} \right)^{-1}
=  \beta    {g^2 \over \bar{g}^2} .
\end{align}
In what follows, we use $\beta$ to denote the initial value, $\bar\beta:=\beta_R$ the running parameter and $\beta_{*}$ the fixed point of RG. 
As $\mu \rightarrow \infty$ or $\bar{g} \rightarrow 0$, $\bar{\beta} \rightarrow + \infty$ for $\beta>0$ and 
$\bar{\beta} \rightarrow -\infty$ for $\beta<0$, while $\bar{\beta}\equiv 0$ for $\beta=0$.
As $\mu \to 0$ or $\bar{g} \rightarrow \infty$, $\bar \beta \to 0$. Hence, $\beta_*=0$ is the IR fixed point for $\beta$.

In the similar way, (\ref{MAalpha}) is cast into 
\begin{align}
 {44 \over 3}    {\partial \over \partial \ln \ln {\mu \over \Lambda_{\rm QCD}}} \alpha_{\rm R}
 =
    -2\alpha_{\rm R}^2+\frac83\alpha_{\rm R}-6
    < 0  .
\label{MAalpha2}
\end{align}
Before solving this equation, we can observe that $\bar \alpha$ is monotonically increasing (decreasing) in decreasing (increasing) $\mu$ towards the IR (UV) direction,
and that there is no fixed point for $\alpha$, in sharp contrast to the Lorentz gauge. 
See Fig.~\ref{fig:GaugeParameterFlowMA}.

The equation (\ref{MAalpha2}) is solved by the integration,
\begin{align}
   \int_{\alpha}^{\bar{\alpha}} {d\alpha_{\rm R} \over -2\alpha_{\rm R}^2+\frac83\alpha_{\rm R}-6}
 = {3 \over 44} \int_{\mu_0}^{\mu} d \ln \ln {\mu \over \Lambda_{\rm QCD}}  .
\label{MAalpha3}
\end{align}
First, we consider sufficiently small $\alpha $ ($|\alpha | \ll 1$), neglecting the order $\alpha^2$ term: 
\begin{align}
   \int_{\alpha}^{\bar{\alpha}} {d\alpha_{\rm R} \over  \alpha_{\rm R}-\frac94 }
 = \frac{2}{11} \int_{\mu_0}^{\mu} d \ln \ln {\mu \over \Lambda_{\rm QCD}}  ,
\end{align}
which yields
\begin{align}
   \bar{\alpha}(\mu) =& {9 \over 4} + \left( \alpha - {9 \over 4} \right) 
\left[ {\ln (\mu/\Lambda_{\rm QCD}) \over \ln (\mu_0/\Lambda_{\rm QCD}) } \right]^{{2 \over 11}}  \nonumber\\
=& {9 \over 4} + \left( \alpha - {9 \over 4} \right)  \left( \frac{g^2}{\bar{g}^2} \right)^{ {2 \over 11} } 
\rightarrow -\infty \quad (\mu \rightarrow \infty, \bar g \to 0) .
\end{align}

Next, we take into account the $O(\alpha^2)$ term too. 
Applying the formula:
\begin{align}
  \int {dx \over ax^2+bx+c} =& {2 \over \sqrt{4ac-b^2}} \arctan {2ax+b \over \sqrt{4ac-b^2}} \nonumber\\&
 (b^2<4ac) ,
\end{align}
to (\ref{MAalpha3}), 
we obtain
%\begin{align}
%  {3 \over 2\sqrt{23}} \arctan {-3\bar{\alpha}+2 \over \sqrt{23}} 
%-  {3 \over 2\sqrt{23}} \arctan {-3\alpha+2 \over \sqrt{23}} 
%= {3 \over 44} \ln \left[ {\ln (\mu/\Lambda_{\rm QCD})  \over \ln (\mu_0/\Lambda_{\rm QCD}) } \right] .
%\end{align}
%or
\begin{align}
   \arctan {-3\bar{\alpha}+2 \over \sqrt{23}} 
=& \arctan {-3\alpha+2 \over \sqrt{23}} \nonumber\\&
+ {\sqrt{23} \over 22} \ln \left[ {\ln (\mu/\Lambda_{\rm QCD})  \over \ln (\mu_0/\Lambda_{\rm QCD}) } \right] .
\end{align}
Thus the running gauge parameter obeys 
\begin{align}
 \bar{\alpha}(\mu) 
%=& {2 \over 3} - {\sqrt{23} \over 3} 
% \tan \left\{ \arctan {-3\alpha+2 \over \sqrt{23}} 
%+ {\sqrt{23} \over 22} \ln \left[ {\ln (\mu/\Lambda_{\rm QCD})  \over \ln (\mu_0/\Lambda_{\rm QCD}) } \right] \right\} 
%\\
=& {2 \over 3} - {\sqrt{23} \over 3} 
 \tan \left\{ \arctan {-3\alpha+2 \over \sqrt{23}} 
+ {\sqrt{23} \over 22} \ln \left[ {g^2  \over  \bar{g}^2} \right] \right\} .
\end{align}
This shows that $\bar{\alpha} \rightarrow -\infty$ as $\mu \rightarrow \infty$ ($\bar g \to 0$) irrespective of the value of $\alpha$. 
Note that $\arctan x$ is multivalued, unless $-\pi/2<\arctan x<\pi/2$. 
%\footnote{
%By making use of the formula
%\begin{align}
% \arctan x \pm \arctan y = \arctan {x\pm y \over 1\mp xy} 
%\quad (|\arctan x + \arctan y| \le \pi/2)
%\end{align}
%}

For higher loop calculations in the MA gauge, see \cite{BG13} .

%%%%%%%%%%   REFERENCES   %%%%%%%%%%

\end{document}